\documentclass[10pt,journal]{IEEEtran}
\usepackage{latexsym,amsmath,amssymb,array,cite,ifthen}

\newcommand{\iftwocolumn}[2]{\ifthenelse{\boolean{@twocolumn}}{#1}{#2}}

\newcommand{\ignore}[1]{}

\newcommand{\Fig}[1]{Fig.~\ref{#1}}

\newcommand{\eqdef}{\stackrel{\scriptscriptstyle\bigtriangleup}{=} }

\newcommand{\vsp}{\rule{0em}{2ex}}    
\newcommand{\andor}{\,/\,}
\newcommand{\cond}{\hspace{0.02em}|\hspace{0.08em}}
\newcommand{\smpl}[2]{#1^{(#2)}}
\newcommand{\vectsubscript}[2]{%
{\!\tiny\left(\!\!\!\begin{array}{c} #1\\ #2 \end{array}\!\!\!\right)}}

\newcommand{\R}{\mathbb{R}}

\newcommand{\calN}{\mathcal{N}}

\newcommand{\tr}{\operatorname{tr}}
\newcommand{\diag}{\operatorname{diag}}
\newcommand{\rvect}{\operatorname{rvect}}
\newcommand{\cvect}{\operatorname{cvect}}

\newcommand{\E}{\operatorname{E}}
\newcommand{\EE}[1]{\mathrm{E}\!\left[{#1}\right]}

\newcommand{\argmax}{\operatornamewithlimits{argmax}}

\newcommand{\msgf}[2]{\protect\overrightarrow{#1}_{\!#2}}
\newcommand{\msgb}[2]{\protect\overleftarrow{#1}_{\!#2}}

\newcommand{\fA}{f_{\mathrm{A}}}
\newcommand{\fB}{f_{\mathrm{B}}}
\newcommand{\fC}{f_{\mathrm{C}}}
\newcommand{\fD}{f_{\mathrm{D}}}
\newcommand{\pB}{p_{\mathrm{B}}}

\newlength{\messagetablewidth}
\newcommand{\tablebox}[1]{\framebox[\messagetablewidth]{%
\begin{minipage}{0.9\messagetablewidth}#1\end{minipage}}\vspace{-0.4pt}}

\newcounter{saveequationcntr}%

\newenvironment{eqntable}%
{\begin{table}%
\setcounter{saveequationcntr}{\value{equation}}%
\setcounter{equation}{0}%
\renewcommand{\theequation}{\Roman{table}.\arabic{equation}}%
}%
{\setcounter{equation}{\value{saveequationcntr}}%
\end{table}}

\newcounter{examplecntr}
{\begin{trivlist}\small\item[]\refstepcounter{examplecntr}%
 {\bfseries Example~\theexamplecntr%
  \ifthenelse{\equal{#1}{}}{}{ (#1)}.
}}%
{\end{trivlist}}

\newcounter{theoremcntr}
{\begin{trivlist}\item[]\refstepcounter{theoremcntr}%
{\bfseries Theorem~\thetheoremcntr%
  \ifthenelse{\equal{#1}{}}{}{ (#1)}.
}}%
{\hfill$\Box$\end{trivlist}}

\newcommand{\eproofnegspace}{\\[-1.5\baselineskip]\rule{0em}{0ex}}

\newcommand{\cent}[1]{\makebox(0,0){#1}}
\newcommand{\pos}[2]{\makebox(0,0)[#1]{#2}}

\begin{document}

\title{Expectation Maximization\\ as Message Passing---%
Part~I:\\ Principles and Gaussian Messages%
{\\[0.5ex] \large October~15, 2009}%
}

\author{%
Justin Dauwels,
Andrew Eckford,
Sascha Korl,
and Hans-Andrea Loeliger%
\thanks{%
J.~Dauwels is with
the Laboratory of Information and Decision Systems 
of the Massachusetts Institute of Technology (MIT), Cambridge, USA.
Email: \texttt{jdauwels@mit.edu}.
}%
\thanks{%
A.~W.~Eckford is with the Dept.\ of Computer Science and Engineering,
York University, 4700 Keele Street, Toronto, ON, Canada M3J 1P3.
Email: \texttt{aeckford@yorku.ca}.
}%
\thanks{%
S.~Korl is with 
Hilti Corporation, 9494 Schaan, Liechtenstein. 
Email: \texttt{sascha.korl@hilti.com}.
}%
\thanks{%
H.-A.~Loeliger is with 
the Dept.\ of Information Technology and Electrical Engineering, 
ETH Zurich, CH-8092 Z\"urich, Switzerland. 
Email: \texttt{loeliger@isi.ee.ethz.ch}.
}%
\thanks{%
Some parts of this work were presented in preliminary form in
\cite{EcPa:imdgm2000c}, \cite{Eck:ceem2004c}, \cite{DKL:ISIT2005c}. 
}%
}

\maketitle

\begin{abstract}
It is shown how expectation maximization (EM) may be viewed 
as a message passing algorithm in factor graphs. 
In particular, a general EM message computation rule is identified.
As a factor graph tool, EM may be used to break cycles in a factor graph, 
and tractable messages may in some cases be obtained 
where the sum-product messages are unwieldy.

As an exemplary application,
the paper considers linear Gaussian state space models. 
Unknown coefficients in such models give rise to multipliers in the corresponding factor graph.
A~main attraction of EM in such cases is 
that it results in purely Gaussian message passing algorithms. 
These Gaussian EM messages are tabulated for several 
(scalar, vector, matrix) multipliers that frequently appear in applications.
\end{abstract}



\section{Introduction}
\label{sec:Introduction}

Graphical models \cite{JoSe:gm} in general
and factor graphs \cite{FKLW:fga1997c,KFL:fg2000,Lg:ifg2004,LDHKLK:fgsp2007}
in particular 
provide a notation for structured system models
that helps to describe and to develop algorithms for detection and
estimation problems.
A large variety of algorithms can be viewed as message passing
algorithms that operate by passing locally computed ``messages'' 
along the edges of the factor graph. 

Expectation maximization (EM) \cite{BaumWelch:1995,DLR:EM1977,McLKr:EME,StSe:cmem2004}
is an iterative technique for parameter estimation 
which is widely used in statistics and signal processing.
EM is a standard tool 
for parameter estimation 
in graphical models \cite{Lau:EMgm1995,Gha:ul2004},
but EM has not traditionally been viewed as a message passing algorithm. 
Examples in communications include 
turbo synchronization \cite{HRVM:EMsynch2003c,NHDLSMLV:turboEMSynch2003c,HVV:isspaem2007},
joint channel estimation and symbol detection \cite{WMK:grjces2004c,CWK:rjce2005,Wu:fgEM2008c},
and distributed source coding \cite{ZRS:2007c}. 

An explicit formulation of a ``factor graph EM algorithm'' 
was proposed in \cite{EcPa:imdgm2000c} and \cite{Eck:ceem2004c},
and a full description of EM as a message passing algorithm
with a general local message computation rule
was presented in \cite{DKL:ISIT2005c},
which is the basis of the present paper.
A similar approach was also pursued by O'Sullivan \cite{Su:mpem2005c} 
and by Herzet et al.~\cite{HVV:isspaem2007}.

In a parallel development, Winn and Bishop made the important observation
that variational inference can be put into message passing form \cite{Wi:thesis2004,WiBi:vmp2005},
and similar observations were made also in \cite{XJR:rmfvar2003} and \cite{NiPa:rctvifg2004}.
In fact, EM message passing may be viewed as 
a special case of variational message passing \cite{Da:vmpfg2007c}. 
However, EM is not specifically addressed (and not even mentioned) 
in \cite{Wi:thesis2004,WiBi:vmp2005,XJR:rmfvar2003}.

In this paper and its companion paper \cite{DEKL:EMId},
we develop the EM algorithm as a general message passing technique for factor graphs.
This formulation may be helpful in several different ways:
\begin{itemize}
\item
EM may be used to estimate unknown parameters in a factor graph model.
\item
EM may be used to break cycles in a factor graph. 
\item
The EM messages are tractable expressions in some cases 
where the sum-product and max-product
message computation rules yields intractable expressions.
\item
Tabulated EM messages for frequently occuring nodes\andor{}factors 
allow the composition of nontrivial EM algorithms  
without additional computations or derivations.
\end{itemize}
Conversely, the flexibility of the factor graph approach 
suggests many variations and extensions of the EM algorithm itself,
as will be discussed in Section~\ref{sec:Conclusions} and in~\cite{DEKL:EMId}.
Moreover, the EM message passing algorithm 
may be seamlessly combined with sum-product and max-product message passing 
in various ways. 

This paper begins with a brief review of standard EM in Section~\ref{sec:EM}
and a detailed development of message passing EM 
in Section~\ref{sec:MessPassEM}.
As quite some time has passed since the publication of 
\cite{EcPa:imdgm2000c,Eck:ceem2004c,DKL:ISIT2005c},
this part of the paper is perhaps mainly tutorial. 

In Section~\ref{sec:Examples}, 
we illustrate 
message passing EM by its application to linear Gaussian models 
(in particular, FIR filters and autoregressive filters) 
with unknown coefficients. 
In these examples, the EM messages 
turn out to be Gaussian, which yields a fully Gaussian algorithm 
for these nonlinear problems. 

These examples also illustrate the use of tabulated EM message computation rules. 
The derivation of the EM message for a particular application is often not trivial 
and tables of precomputed EM messages can therefore be helpful. 
In Section~\ref{sec:GaussianMultiplierEM}, 
we present tables of EM messages out of various ``multipliers'' that 
arise naturally in linear Gaussian models with unknown coefficients.

The proofs of these tabulated message computation rules are given 
in Appendices \ref{appsec:ProofsTableGaussMultEM}--\ref{appsec:VXY}.
Appendices \ref{appsec:ProofsTableVxmx} and~\ref{appsec:VXY} 
rely on Gaussian sum-product messages tabulated in \cite{LDHKLK:fgsp2007}, 
which further illustrates the use of tabulated message computation rules.

Some concluding remarks are offered in Section~\ref{sec:Conclusions}. 

The companion paper \cite{DEKL:EMId} begins with discrete variables
and makes a tour through EM algorithms ranging from hidden Markov models 
to independent factor analysis.

In this paper, we will use Forney-style factor graphs 
(also called normal factor graphs) as in
\cite{Lg:ifg2004} and~\cite{LDHKLK:fgsp2007}, 
a variation due to Forney \cite{Forney:nr2002}
of factor graphs as in \cite{KFL:fg2000}.
The reader is specifically referred to \cite{LDHKLK:fgsp2007} for details 
of the factor graph notation.
In particular,
we will use arrows (as in $\msgf{\mu}{}$ and $\msgb{\mu}{}$) for sum-product messages, 
and we will use capital letters for unknown variables 
(i.e., functions of the configuration space)
and lower-case letters for particular values of a variable. 

%
From Section~\ref{sec:Examples} onward, 
multivariate Gaussian distributions
will be prominent.
Such distributions will be parameterized either 
by a mean vector $m$ and a covariance matrix $V$
or by the inverse covariance matrix (``weight matrix'') $W=V^{-1}$ 
and the transformed mean vector $Wm$.
For Gaussian messages, these parameters will be denoted by 
$\msgf{m}{}$, $\msgf{V}{}$, etc., 
as in \cite{LDHKLK:fgsp2007}.
We will sometimes allow messages to be degenerate (non-integrable) ``Gaussians''
$e^{-\frac{1}{2}(x^TWx-2x^TWm)}$ 
where the weight matrix $W$ is positive semi-definite and singular 
rather than positive definite.

\section{Review of the EM Algorithm}
\label{sec:EM}

We begin by reviewing the EM algorithm in a setting which 
is suitable for the purpose of this paper.
Suppose we wish to find
\begin{equation} \label{eqn:thetamax}
\hat\theta_{\mathrm{max}} \eqdef \argmax_{\theta} f(\theta)
\end{equation}
for some function $f:\R^n\rightarrow \R$. 
We assume that $f(\theta)$ is the ``marginal'' of some real-valued
function $f(x,\theta)$, i.e., 
\begin{equation} \label{eqn:fasMarginal}
f(\theta) = \int_x f(x,\theta)\, dx
\end{equation}
where $\int_x g(x)\, dx$ denotes integration of
$g(x)$ over the whole range of $x$. 
(The integral in (\ref{eqn:fasMarginal}) may be replaced 
by a sum if $x$ is discrete, with obvious corresponding changes in subsequent expressions.)
The function $f(x,\theta)$ is
assumed to be nonnegative:
\begin{equation}
f(x,\theta) \geq 0 \text{~~~~for all $x$ and all $\theta$}.
\end{equation}
In addition, we assume $0<f(\theta)<\infty$ for all $\theta$. 
In other words, for any fixed $\theta$, 
$f(x,\theta) / f(\theta)$ is a probability density over~$x$.
We will also assume that the integral 
$\int_x f(x,\theta) \log f(x,\theta')\, dx$ exists for all $\theta$, $\theta'$.

The EM algorithm attempts to compute (\ref{eqn:thetamax}) as
follows:
\begin{enumerate}
\item
Make some initial guess $\smpl{\hat\theta}{0}$.
\item
Expectation step: evaluate
\begin{equation} \label{eqn:EMEstep}
\smpl{f}{k}(\theta) \eqdef \int_x f(x,\smpl{\hat\theta}{k}) \log
f(x,\theta)\, dx.
\end{equation}
(The base of the logarithm is immaterial.)
\item
Maximization step: compute
\begin{equation} \label{eqn:EMmaxstep}
\smpl{\hat\theta}{k+1} \eqdef \argmax_{\theta} \smpl{f}{k}(\theta).
\end{equation}
\item
Repeat 2--3 until convergence or until the available time is over.
\end{enumerate}
The main property of the EM algorithm is 
\begin{equation} \label{eqn:EMTheorem}
f(\smpl{\hat\theta}{k+1}) \geq f(\smpl{\hat\theta}{k}).
\end{equation}
For the reader's convenience, a concise proof of~(\ref{eqn:EMTheorem}) 
is given in Appendix~\ref{appsec:ProofEM}.
In many applications, 
the expressions (\ref{eqn:EMEstep}) and (\ref{eqn:EMmaxstep}) 
turn out to be quite manageable 
and simpler than the direct maximization~(\ref{eqn:thetamax}).

In typical applications, $f(x,\theta)$ is extended to $f(x,y,\theta)$,
where $y$ is known and fixed. 
The function $f(x,y,\theta)$ 
is either a probability density over $x$ and $y$ 
with parameter $\theta$ 
or it is a joint probability density over $x$, $y$, and $\theta$.
In the EM literature, 
$y$ is called the observed data, 
$x$ is called the missing (unobserved) data,
and the pair $(x,y)$ is called the complete data.

\section{EM as a Message Passing Algorithm}
\label{sec:MessPassEM}

We now consider EM in factor graphs. 
We will do this in several steps. 
The resulting message passing algorithm will be summarized 
in Section~\ref{sec:MessPassEMSummary}.

We henceforth assume that all logarithms are natural logarithms.

\subsection{Trivial Factor Graph}

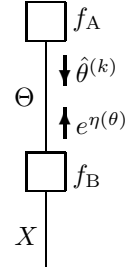
\begin{figure}
\begin{center}
\begin{picture}(5,35)(0,0)
\put(0,30){\framebox(5,5){}}  \put(6,32){$\fA$}
\put(2.5,15){\line(0,1){15}}  \put(1,22.5){\pos{r}{$\Theta$}}
 {\thicklines
  \put(5,28){\vector(0,-1){4}} \put(6.5,24.5){$\smpl{\hat\theta}{k}$}
  \put(5,17){\vector(0,1){4}}  \put(6.5,17.5){$e^{\eta(\theta)}$}
 }
\put(0,10){\framebox(5,5){}}  \put(6,11){$\fB$}
\put(2.5,0){\line(0,1){10}}   \put(1.5,4){\pos{r}{$X$}}
\end{picture}
\caption{\label{fig:TrivialFactorGraph}%
Factor graph of (\ref{eqn:TrivialFactorization})
with EM~message $e^{\eta(\theta)}$.}
\end{center}
\end{figure}

We first consider a trivial factorization 
\begin{equation} \label{eqn:TrivialFactorization}
f(x,\theta) = \fA(\theta) \fB(x,\theta),
\end{equation}
the factor graph of which is shown in \Fig{fig:TrivialFactorGraph}.
(In typical applications, $\fA(\theta)$ is either a prior probability or constant.)
In this setup, the EM algorithm amounts to iterative computation
of a downward message $\smpl{\hat\theta}{k}$ and an upward message $e^{\eta(\theta)}$ 
as follows.
\begin{trivlist}
\item
\emph{Upward message (EM~message):} $e^{\eta(\theta)}$ with
\begin{IEEEeqnarray}{rCl}
\eta(\theta) & \eqdef &
   \frac{\int_x \fB(x,\smpl{\hat\theta}{k}) \log\fB(x,\theta)\, dx}
        {\int_x \fB(x,\smpl{\hat\theta}{k})\, dx}
        \IEEEeqnarraynumspace
   \label{eqn:UpwardsMessageTrivialGraph} \\
& = &  \E_{\pB}\!\left[ \log \fB(X,\theta) \right],
          \label{eqn:UpwardsMessageTrivialGraphExpectation}
\end{IEEEeqnarray}
where $\E_{\pB}$ denotes the expectation with respect to the
probability distribution
\begin{equation}
    \label{eqn:normalized-probability}
\pB(x \cond \smpl{\hat\theta}{k}) \eqdef
  \frac{\fB(x,\smpl{\hat\theta}{k})}
       {\int_{x'}\fB(x',\smpl{\hat\theta}{k})\, dx'}
\end{equation}
\item
\emph{Downward message:}
\begin{IEEEeqnarray}{rCl}
\smpl{\hat\theta}{k+1} &=&  \argmax_\theta \left( \fA(\theta)
\cdot e^{\eta(\theta)} \right)
          \label{eqn:DownwardsMessageTrivialGraphMultiplicative} \\
&=&  \argmax_\theta \left( \log \fA(\theta) + \eta(\theta)
\right).
          \label{eqn:DownwardsMessageTrivialGraph}
\end{IEEEeqnarray}
\end{trivlist}

\noindent The equivalence of this message passing algorithm with
(\ref{eqn:EMEstep}) and (\ref{eqn:EMmaxstep}) may be seen as
follows. From (\ref{eqn:EMEstep}) and (\ref{eqn:EMmaxstep}), we have
\iftwocolumn{
 \begin{IEEEeqnarray}{rCl}
 \IEEEeqnarraymulticol{3}{l}{
 \smpl{\hat\theta}{k+1}
 }\nonumber\\\quad
}%
{\begin{IEEEeqnarray}{rCl}
 \smpl{\hat\theta}{k+1}
}
&=& \argmax_{\theta} \int_x f(x,\smpl{\hat\theta}{k}) \log f(x,\theta)\, dx \IEEEeqnarraynumspace\\
&=& \argmax_\theta 
    \int_x \fA(\smpl{\hat\theta}{k})
           \fB(x,\smpl{\hat\theta}{k}) 
     \iftwocolumn{\nonumber\\ && {~~~~~~~~~~~~~~~~} \cdot}{}
     \log\!\big( \fA(\theta) \fB(x,\theta) \big)\, dx \IEEEeqnarraynumspace\\
&=& \argmax_\theta 
    \int_x \fB(x,\smpl{\hat\theta}{k})  
    \iftwocolumn{\nonumber\\ && {~~~~~~~~} \cdot }{}
    \Big( \log \fA(\theta) + \log \fB(x,\theta) \Big)\, dx \IEEEeqnarraynumspace\\
&=& \argmax_\theta \Bigg( \log \fA(\theta) 
      \iftwocolumn{\nonumber\\ && {~~~~} }{}
        + \frac{\int_x \fB(x,\smpl{\hat\theta}{k}) \log \fB(x,\theta)\, dx}
                {\int_{x'} \fB(x',\smpl{\hat\theta}{k})\, dx'}
         \,\Bigg) \IEEEeqnarraynumspace
\end{IEEEeqnarray}
which is equivalent to (\ref{eqn:UpwardsMessageTrivialGraph}) and
(\ref{eqn:DownwardsMessageTrivialGraph}).

Some remarks:
\begin{enumerate}
\item
The quantity $\eta(\theta)$ may be viewed as a ``log-domain''
summary of $f_B$. 
The corresponding ``probability domain'' summary $e^{\eta(\theta)}$ is
consistent with the semantics of factor graphs where 
messages are ``summaries'' of factors
(cf.~(\ref{eqn:DownwardsMessageTrivialGraphMultiplicative}) 
and~(\ref{eqn:UpwardsEMMessageFactorized})). 
We will refer to $e^{\eta(\theta)}$ as the \emph{EM message}.
\item
\label{enum:TrivFFGRemarkConstInh}
A constant may be added to $\eta(\theta)$ without
affecting (\ref{eqn:DownwardsMessageTrivialGraph}).
\item
If $\fA(\theta)$ is constant, the normalization in
(\ref{eqn:UpwardsMessageTrivialGraph}) can be omitted. More
generally, the normalization in
(\ref{eqn:UpwardsMessageTrivialGraph}) can be omitted if
$\fA(\theta)$ is constant for all $\theta$ such that
\mbox{$\fA(\theta)\neq 0$}
(i.e., if $\fA(\theta)$ expresses a constraint); 
this case occurs in many applications.
\item
Nothing changes if we introduce a known observation (i.e., a
constant argument) $y$ into $f$ such that
(\ref{eqn:TrivialFactorization}) becomes 
$f(x,y,\theta) = \fA(y,\theta) \fB(x,y,\theta)$. 
\end{enumerate}

\subsection{Nontrivial Factor Graph}
\label{sec:NontrivialFactorGraphEM}

We now come to the heart of the matter: 
if $\theta$ is a vector, $\theta=(\theta_1,\theta_2,\ldots)$, 
and if 
$\fB$
has a nontrivial factor graph, then the EM~message $e^{\eta(\theta)}$ 
splits into messages $e^{\eta_1(\theta_1)}$, $e^{\eta_2(\theta_2)}$,~\ldots
that can be computed ``locally'' in the factor graph of $\fB$.

To see this, consider the following example 
(which actually covers the general case). 
Let $\theta=(\theta_1, \theta_2)$, 
let $x=(x_1,x_2,x_3)$, and let 
\begin{equation} \label{eqn:TwoFactors}
\fB(x,\theta) = \fC(x_1,x_2,\theta_1) \fD(x_2,x_3,\theta_2),
\end{equation}
the factor graph of which is shown in \Fig{fig:TwoFactorsEM}. 
\begin{figure}
\begin{center}
\begin{picture}(60,42)(0,0)
%
\put(15,35){\framebox(30,7){}}   \put(13.5,38.5){\pos{r}{$\fA$}}
\put(17.5,15){\line(0,1){20}}    \put(16.5,26){\pos{r}{$\Theta_1$}}
 {\thicklines
  \put(20,33){\vector(0,-1){4}}  \put(22,30.5){\pos{l}{$\smpl{\hat\theta_1}{k}$}}
  \put(20,20){\vector(0,1){4}}   \put(22,23){\pos{l}{$e^{\eta_1(\theta_1)}$}}
 }
\put(42.5,15){\line(0,1){20}}    \put(41.5,26){\pos{r}{$\Theta_2$}}
 {\thicklines
  \put(45,33){\vector(0,-1){4}}  \put(47,30.5){\pos{l}{$\smpl{\hat\theta_2}{k}$}}
  \put(45,20){\vector(0,1){4}}   \put(47,23){\pos{l}{$e^{\eta_2(\theta_2)}$}}
 }
\put(0,12.5){\line(1,0){15}}     \put(3,15.5){\cent{$X_1$}}
\put(15,10){\framebox(5,5){}}    \put(17.5,6){\cent{$\fC$}}
\put(20,12.5){\line(1,0){20}}    \put(30,9.5){\cent{$X_2$}}
\put(40,10){\framebox(5,5){}}    \put(42.5,6){\cent{$\fD$}}
\put(45,12.5){\line(1,0){15}}    \put(57,15.5){\cent{$X_3$}}
\put(10,2){\dashbox(40,16){}}    \put(51.5,5){\pos{l}{$\fB$}}
\end{picture}
\caption{\label{fig:TwoFactorsEM}%
Factor graph of (\ref{eqn:TwoFactors}), a refinement of \Fig{fig:TrivialFactorGraph}.
}
\end{center}
\end{figure}
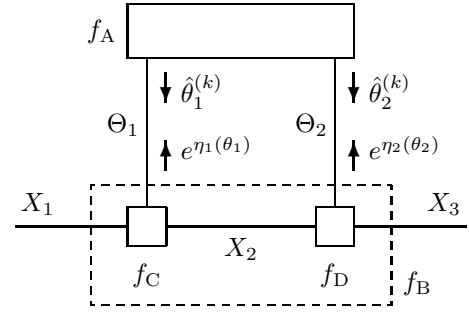
In this case, (\ref{eqn:UpwardsMessageTrivialGraphExpectation})
splits into
\begin{IEEEeqnarray}{rCl}
\eta(\theta_1,\theta_2) 
&=& 
  \E_{\pB}\!\!\left[ \log \Big( \fC(X_1,X_2,\theta_1) \fD(X_2,X_3,\theta_2) \Big) \right] 
  \IEEEeqnarraynumspace\\
&=&  \eta_1(\theta_1) + \eta_2(\theta_2)
     \label{eqn:TwoFactorsSplitExpectation}
\end{IEEEeqnarray}
with
\begin{equation} \label{eqn:TwoFactorsElog1}
\eta_1(\theta_1) \eqdef \E_{\pB}\!\left[ \log \fC(X_1,X_2,\theta_1) \right]
\end{equation}
and
\begin{equation} \label{eqn:TwoFactorsElog2}
\eta_2(\theta_2) \eqdef \E_{\pB}\!\left[ \log \fD(X_2,X_3,\theta_2) \right].
\end{equation}
The EM~message 
$e^{\eta(\theta)}$ thus factors as
\begin{equation} \label{eqn:UpwardsEMMessageFactorized}
e^{\eta(\theta_1,\theta_2)} = e^{\eta_1(\theta_1)} e^{\eta_2(\theta_2)},
\end{equation}
and the factors $e^{\eta_1(\theta_1)}$ and $e^{\eta_2(\theta_2)}$
may be viewed as upward messages along the edge 
$\Theta_1$ and $\Theta_2$, respectively, in the factor graph 
of \Fig{fig:TwoFactorsEM}.
The downward messages in \Fig{fig:TwoFactorsEM} 
are the estimates 
\begin{equation}  \label{eqn:EMMaxMessages}
(\smpl{\hat{\theta}_1}{k+1},\smpl{\hat{\theta}_2}{k+1})
= \argmax_{(\theta_1,\, \theta_2)} \fA(\theta_1, \theta_2) 
   e^{\eta_1(\theta_1)} e^{\eta_2(\theta_2)}
\end{equation}
as is obvious from (\ref{eqn:DownwardsMessageTrivialGraphMultiplicative}) 
and~(\ref{eqn:UpwardsEMMessageFactorized}).

The expectation in (\ref{eqn:TwoFactorsElog1}) 
may be computed with respect to the 
probability distribution 
\begin{equation} \label{eqn:TwoFactorsLocalProb1}
\pB(x_1,x_2 \cond \smpl{\hat{\theta}}{k}) 
 \eqdef \int_{x_3} \pB(x_1,x_2,x_3 \cond \smpl{\hat{\theta}}{k})\, dx_3,
\end{equation}
which is the marginal of $\pB$ 
with respect to the arguments of~$\fC$,
and the expectation in (\ref{eqn:TwoFactorsElog2}) 
may be computed with respect to the 
probability distribution
\begin{equation} \label{eqn:TwoFactorsLocalProb2}
\pB(x_2,x_3 \cond \smpl{\hat{\theta}}{k}) 
 \eqdef \int_{x_1} \pB(x_1,x_2,x_3 \cond \smpl{\hat{\theta}}{k})\, dx_1,
\end{equation}
which is the marginal of $\pB$ 
with respect to the arguments of~$\fD$.

Going through this derivation, we note that the generalization 
to an arbitrary factor graph for $\fB$ is immediate. 
Note, in particular, that the splitting of the expectation in 
(\ref{eqn:TwoFactorsSplitExpectation}) 
does not assume that the factor graph of $\fB$ is cycle-free.
If $g(x_1,\ldots,x_m,\theta_g)$ 
is a generic node\andor{}factor in the factor graph of $\fB$,
we obtain $\eta_g(\theta_g)$ 
as in (\ref{eqn:GenLocalEMRuleExpect}) and~(\ref{eqn:GenLocalEMRuleIntegral})
in Table~\ref{tab:ElogMessageGeneral}
with
\begin{IEEEeqnarray}{rCl}
p_\text{local}(x_1,\ldots,x_m \cond \hat\theta)
 & \eqdef & \int_{x: x_1\ldots x_m \text{fixed}} \pB(x \cond\hat\theta)\, dx \\
 & \propto & \int_{x: x_1\ldots x_m \text{fixed}} \fB(x,\hat\theta)\, dx, 
    \IEEEeqnarraynumspace\label{eqn:LocalMarginal}
\end{IEEEeqnarray}
where ``$\propto$'' denotes equality up to a scale factor.
Note that the missing scale factor in (\ref{eqn:LocalMarginal})
can be locally recovered by integrating (\ref{eqn:LocalMarginal}) over $x_1\ldots x_m$.
It remains to make the step from (\ref{eqn:LocalMarginal}) 
to (\ref{eqn:LocalProbBySumProduct}) in Table~\ref{tab:ElogMessageGeneral}.

\subsection{Using Sum-Product Message Passing for the Local Expectations}

If the factor graph of $\fB(x,\hat\theta)$  
is cycle-free (after removing the edges for $\Theta=\hat\theta$), 
then the marginals~(\ref{eqn:LocalMarginal}) can be computed 
by sum-product message passing (see \cite{Lg:ifg2004,LDHKLK:fgsp2007}) in this factor graph.
As above, 
let $g(x_1,\ldots,x_m,\theta_g)$ be a generic node\andor{}factor in the factor graph of $\fB$.  
Then (\ref{eqn:LocalMarginal}) may be computed 
as in (\ref{eqn:LocalProbBySumProduct}) in Table~\ref{tab:ElogMessageGeneral}, 
where $\msgf{\mu}{X_\ell}$ denotes the incoming sum-product message 
along the variable\andor{}edge $X_\ell$
computed for $\Theta=\hat\theta$.

\begin{eqntable}
\caption{\label{tab:ElogMessageGeneral}%
EM message $e^{\eta_g(\theta_g)}$ out of a generic node\andor{}factor $g$.}
\begin{center}
\tablebox{
\vspace{5mm}
\begin{center}
\begin{picture}(35,35)(0,0)
%
\put(15,20){\line(0,1){15}}      \put(13.5,27){\pos{r}{$\Theta_g$}}
 {\thicklines
  \put(17,33){\vector(0,-1){4}}  \put(18.5,31.5){\pos{l}{$\hat\theta_g$}}
  \put(17,22){\vector(0,1){4}}   \put(18.5,25){\pos{l}{$e^{\eta_g(\theta_g)}$}}
 }
\put(0,10){\framebox(35,10){}}   \put(-1.5,15){\pos{r}{$g$}}
\put(5,0){\line(0,1){10}}        \put(4,3){\pos{r}{$X_1$}}
 {\thicklines
  \put(7,1){\vector(0,1){4}}     \put(8.5,0){$\msgf{\mu}{X_1}$}
 }
 \put(17.5,7){\cent{$\cdots$}}
\put(30,0){\line(0,1){10}}       \put(29,3){\pos{r}{$X_m$}}
 {\thicklines
  \put(32,1){\vector(0,1){4}}    \put(33.5,0){$\msgf{\mu}{X_m}$}
 }
\end{picture}
\end{center}
\vspace{3mm}
\begin{IEEEeqnarray}{rCl}
\eta_g(\theta_g) &=& \E_{p_\text{local}}\!\left[\vsp\log g(X_1,\ldots,X_m,{\theta_g}) \right] 
       \label{eqn:GenLocalEMRuleExpect}\\
&=& \int_{x_1,\ldots,x_m} p_\text{local}(x_1,\ldots,x_m \cond \hat\theta) 
         \nonumber\\ && {~~~~~} \cdot{}
         \log g(x_1,\ldots,x_m,\theta_g)\, dx_1\cdots dx_m
         \IEEEeqnarraynumspace
         \label{eqn:GenLocalEMRuleIntegral}
\end{IEEEeqnarray}
with
\begin{IEEEeqnarray}{rCl}
  \IEEEeqnarraymulticol{3}{l}{
  p_\text{local}(x_1,\ldots,x_m \cond \hat\theta)
  }\nonumber\\\quad
 & \propto & g(x_1,\ldots,x_m, \hat\theta_g)\, 
         \msgf{\mu}{X_1}(x_1)\cdots \msgf{\mu}{X_m}(x_m)
       \IEEEeqnarraynumspace \label{eqn:LocalProbBySumProduct}
\end{IEEEeqnarray}
where $\msgf{\mu}{X_\ell}$ denotes the incoming sum-product message 
along the variable\andor{}edge $X_\ell$
computed for $\Theta=\hat\theta$.
\vspace{1ex}

A constant scale factor $\gamma$ in $g$ results 
in a scale factor $\gamma$ in $e^{\eta_g(\theta_g)}$  
which can be ignored.
\vspace{2.5mm}
}
\end{center}
\end{eqntable}

For example, we can write (\ref{eqn:TwoFactorsLocalProb1}) as
\begin{IEEEeqnarray}{rCl}
\pB(x_1,x_2 \cond \hat\theta)
& \propto & \int_{x_3} \! \fC(x_1,x_2,\hat\theta_1) 
                          \fD(x_2,x_3,\hat\theta_2)\, dx_3
                          \IEEEeqnarraynumspace\\
& = & \fC(x_1,x_2,\hat\theta_1) \msgb{\mu}{X_2}(x_2)
      \label{eqn:plocalSimpleExampleSumProd}
\end{IEEEeqnarray}
where $\msgb{\mu}{X_2}$ is the right-to-left sum-product message 
along the edge $X_2$ computed for $\Theta=\hat\theta$. 
(A constant message $\msgf{\mu}{X_1}(x_1)=1$ may be added 
as a factor in (\ref{eqn:plocalSimpleExampleSumProd}).)

\subsection{Using Max-Product Message Passing for the Maximization}

If $\fA$ can be factored into a cycle-free factor graph,
then the maximization (\ref{eqn:EMMaxMessages}) 
(and its obvious generalization to general factor graphs) 
can be carried out by max-product message passing in the factor graph of $\fA$.
This applies, in particular, to the standard case 
where $\fA(\theta_1,\theta_2,\ldots)$ 
expresses the equality constraint \mbox{$\Theta_1=\Theta_1=\ldots$},
which we will encounter in Section~\ref{sec:Examples}.

\subsection{Putting it Together}
\label{sec:MessPassEMSummary}

\begin{figure}
\begin{center}
\setlength{\unitlength}{0.73mm}
\begin{picture}(118,55)(0,0)
%
\put(0,45){\framebox(110,10){}}  \put(112,45){$\fA$}
\put(0,7){\dashbox(110,18){}}    \put(112,7){$\fB$}
\put(5,15){\framebox(5,5){}}     \put(7.5,11){\cent{$f_0$}}
\put(10,17.5){\vector(1,0){15}}  \put(17.5,20.8){\cent{$X_0$}}
\put(25,15){\framebox(5,5){}}    \put(26,11){\pos{r}{$f_1$}}
\put(27.5,20){\line(0,1){25}}    \put(26.5,35){\pos{r}{$\Theta_1$}}
 {\thicklines
  \put(30,42){\vector(0,-1){5}}  \put(32,38){$\hat\theta_1$}
  \put(30,28){\vector(0,1){5}}   \put(32,28){$e^{\eta_1(\theta_1)}$}
 }
\put(27.5,15){\line(0,-1){15}}   \put(29,0){$y_1$}
\put(30,17.5){\vector(1,0){15}}  \put(37.5,20.8){\cent{$X_1$}}
\put(45,15){\framebox(5,5){}}    \put(46,11){\pos{r}{$f_2$}}
\put(47.5,20){\line(0,1){25}}    
 {\thicklines
  \put(50,42){\vector(0,-1){5}}  \put(51.5,38){$\hat\theta_2$}
  \put(50,28){\vector(0,1){5}}   \put(51.5,28){$e^{\eta_2(\theta_2)}$}
 }
\put(47.5,15){\line(0,-1){15}}   \put(49,0){$y_2$}
\put(50,17.5){\line(1,0){10}}    \put(57.5,20.8){\cent{$X_2$}}
\put(70,35){\cent{$\ldots$}}
\put(70,15){\cent{$\ldots$}}
\put(80,17.5){\vector(1,0){10}}  \put(82.5,20.8){\cent{$X_{n-1}$}}
\put(90,15){\framebox(5,5){}}    \put(91,11){\pos{r}{$f_n$}}
\put(92.5,20){\line(0,1){25}}    \put(91.5,35){\pos{r}{$\Theta_n$}}
 {\thicklines
  \put(95,42){\vector(0,-1){5}}  \put(96.5,38){$\hat\theta_n$}
  \put(95,28){\vector(0,1){5}}   \put(96.5,28){$e^{\eta_n(\theta_n)}$}
 }
\put(92.5,15){\line(0,-1){15}}   \put(94,0){$y_n$}
\put(95,17.5){\vector(1,0){10}}  \put(102.5,21){\cent{$X_n$}}
\end{picture}
\caption{\label{fig:GenStateSpaceEM}%
Application of EM to general state space model.}
\end{center}
\end{figure}
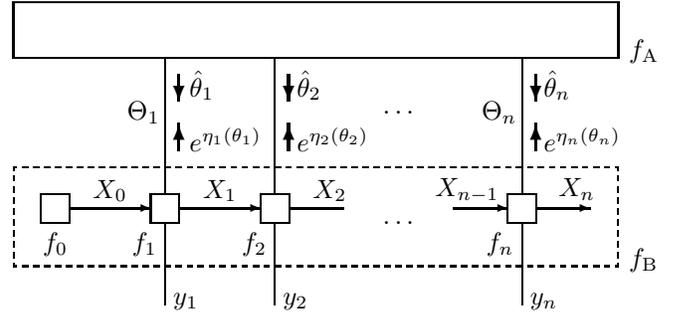

Let us summarize the findings of this section by considering 
the factor graph of \Fig{fig:GenStateSpaceEM}, 
which is an easy generalization of \Fig{fig:TwoFactorsEM}.
Note that removing the edges $\Theta_1,\ldots,\Theta_n$ 
cuts the factor graph (\Fig{fig:GenStateSpaceEM}) into 
two cycle-free components.
Let $\theta \eqdef (\theta_1,\ldots,\theta_n)$, 
$x \eqdef (x_1,\ldots,x_n)$, and $y \eqdef(y_1,\ldots,y_n)$. 
Suppose that we wish to find 
\begin{equation}
\hat{\theta} = \argmax_{\theta} \fA(\theta) \int_x \fB(x,y,\theta)\, dx
\end{equation}
for fixed known $y$.
In this case, the EM algorithm applies as follows:
\begin{enumerate}
\item
Make some initial guess $\hat{\theta} = (\hat{\theta}_1, \ldots, \hat{\theta}_n)$.
\item
Perform forward-backward sum-product message passing through the factor graph of~$\fB$
(with $\hat{\theta}_\ell$ plugged into $f_\ell$ for $\ell=1,\ldots, n$).
\item
Compute the EM messages $e^{\eta_1(\theta_1)}$, \ldots, $e^{\eta_n(\theta_n)}$
as in Table~\ref{tab:ElogMessageGeneral}. 
In this case, 
we obtain
\begin{equation}
\eta_\ell(\theta_\ell) = 
  \E_{p_\text{local}}\!\left[\vsp\log f_\ell(X_{\ell-1},X_\ell,y_\ell,\theta_\ell) \right]
\end{equation}
where the expectation is with respect to the probability density
\begin{IEEEeqnarray}{rCl}
 p_\text{local}(x_{\ell-1}, x_\ell \cond y_\ell, \hat\theta)
 & \propto & f_\ell(x_{\ell-1},x_\ell,y_\ell,\hat{\theta}_\ell) 
   \iftwocolumn{\IEEEeqnarraynumspace\nonumber\\ && \hspace{-0em} \cdot}{}
     \msgf{\mu}{X_{\ell-1}}(x_{\ell-1}) \msgb{\mu}{X_\ell}(x_\ell)
     \IEEEeqnarraynumspace
\end{IEEEeqnarray}
where $\msgf{\mu}{X_{\ell-1}}$ and $\msgb{\mu}{X_\ell}$ denote 
sum-product messages.
\item
Compute new estimates 
\begin{IEEEeqnarray}{rCl}
\hat{\theta} & = & (\hat{\theta}_1, \ldots, \hat{\theta}_n) \\
& = & \argmax_{(\theta_1,\ldots,\theta_n)} \fA(\theta_1,\ldots,\theta_n)\, 
                e^{\eta_1(\theta_1)} \cdots e^{\eta_n(\theta_n)}.
                \IEEEeqnarraynumspace
\end{IEEEeqnarray}
If $\fA$ has a cycle-free factor graph, this maximization 
may be carried out by max-product message passing in that factor graph.
\item
Repeat 2--4 until convergence or until the available time is over. 
\end{enumerate}

All this applies to general factorizations of $\fA$ and $\fB$
provided that the resulting factor graphs (without the edges $\Theta_1$,\ldots,$\Theta_n$) 
are cycle-free.

If the factor graphs of $\fA$ and $\fB$ are not cycle-free, 
the same local computation rules can be used nonetheless 
and seem to work well in some applications,
cf.\ \cite{HRVM:EMsynch2003c,NHDLSMLV:turboEMSynch2003c,HVV:isspaem2007,WMK:grjces2004c,CWK:rjce2005,Wu:fgEM2008c,ZRS:2007c}.

In many cases, the computation of an EM message 
according to Table~\ref{tab:ElogMessageGeneral} 
requires substantial additional work. 
Precomputed tables of such messages 
for frequently occuring nodes\andor{}factors 
can therefore be useful,
as will be demonstrated in Sections \ref{sec:Examples} 
and~\ref{sec:GaussianMultiplierEM}.

\subsection{An Issue: Hard Constraints and Grouping}
\label{sec:IssueHardConstraintsGrouping}

Nodes in factor graphs often express ``hard'' constraints \cite{Lg:ifg2004,LDHKLK:fgsp2007}.
For example, the constraint $X_1=X_2$ (for real variables $X_1$ and $X_2$) 
may be expressed by the node\andor{}factor
\mbox{$\delta(x_1-x_2)$}, where $\delta$ denotes the Dirac delta.
It turns out that the EM message computation rule
of Table~\ref{tab:ElogMessageGeneral} 
should not be applied to such constraint nodes; 
the typical outcome of the attempt will be a degenerate EM message $e^{\eta_s(\theta_s)}$
that expresses the constraint $\Theta_s = \hat{\theta}_s$,
which stalls the EM algorithm.

For example, assume that $X_1, X_2, \Theta$ are real variables 
and the node\andor{}factor
\begin{equation}
g(x_1,x_2,\theta) = \delta(x_1-x_2\theta)
\end{equation}
expresses the constraint $X_1 = X_2\Theta$.
Then
\begin{IEEEeqnarray}{rCl}
\eta(\theta) 
& \propto & 
    \int_{x_1} \int_{x_2} g(x_1,x_2,\hat\theta) \msgf{\mu}{X_1}(x_1) \msgf{\mu}{X_2}(x_2)
    \iftwocolumn{\nonumber\\ && {~~~~~~~~~~~~~~~~~~~~} \cdot}{}
    \log g(x_1,x_2,\theta)\, dx_1 dx_2 \\
& = & \int_{x_2} \msgf{\mu}{X_1}(x_2 \hat\theta) \msgf{\mu}{X_2}(x_2)
      \log g(x_2\hat\theta,x_2,\theta)\, dx_2
      \IEEEeqnarraynumspace \\
& = & \int_{x_2} \msgf{\mu}{X_1}(x_2 \hat\theta) \msgf{\mu}{X_2}(x_2)
      \log \delta(x_2(\hat\theta - \theta))\, dx_2,
      \IEEEeqnarraynumspace
\end{IEEEeqnarray}
which is obviously pathological
and illustrates the issue.

It is usually easy to avoid this problem by 
grouping constraint nodes with 
adjacent ``soft'' factors\andor{}nodes, 
as will be illustrated in Sections~\ref{sec:Examples} and~\ref{sec:GaussianMultiplierEM}.

\section{Examples: Identification of Linear Systems}
\label{sec:Examples}

The following two examples arise in many applications. 
The use of EM to problems of this kind 
is not new, but neither is it trivial
\cite{ShSt:TimeSeriesEM1982,RoGha:LinGaussRev1999,GiNi:rmldynsys2005}. 
In communications, the example of Section~\ref{sec:ExampleFIR}
may arise in channel estimation and the example of Section~\ref{sec:AutoregressiveFilter}
may arise in estimating the parameters of non-white Gaussian noise.

\subsection{FIR Filter Identification with Unknown Input Signal}
\label{sec:ExampleFIR}

Let $X_k\in\R^n$, $k=0,1,2,\ldots, N$,  
be the time-$k$ state of a finite impulse response (FIR) filter 
with random input signal $U_k\in\R$, $k=1,2,\ldots, N$.
Specifically,
\begin{equation} \label{eqn:NextStateFIR}
X_k = A X_{k-1} + b U_k
\end{equation}
with $n\times n$ matrix
\begin{equation}
A = \left( \begin{array}{cc} 0 & 0 \\ I_{n-1} & 0 \end{array} \right)
\end{equation}
(where $I_{n-1}$ is the $(n-1)\times(n-1)$ identity matrix)
and with
\begin{equation}
b = \left( 1, 0, \ldots, 0 \right)^T.
\end{equation}
We assume that the input signal $U_1, U_2, \ldots$ is zero-mean white Gaussian noise 
with variance $\sigma_U^2$.
We observe a noisy scalar output signal
\begin{equation} \label{eqn:NoisyOutputFIR}
Y_k = \Theta^T X_k + Z_k
\end{equation}
where $\Theta$ is an unknown real column vector and
where $Z_k$ is zero-mean white Gaussian noise 
with variance $\sigma_Z^2$.
From the observations $Y_k=y_k$, $k=1,2,\ldots,N$, 
we wish to estimate $\Theta$. 
Specifically, we wish to compute
the maximum-likelihood estimate
\begin{IEEEeqnarray}{rCl}
\hat{\theta} 
& = & \argmax_{\theta} p(y \cond \theta) \\
& = & \argmax_{\theta} \int_{u} \int_{x} \int_{z} p(u,x,y,z \cond \theta)\, dz\, dx\, du,
      \IEEEeqnarraynumspace
\end{IEEEeqnarray}
where $y$ is defined as $y\eqdef (y_1,\ldots,y_N)$ 
and where $u$, $x$, $z$ are defined analogously.

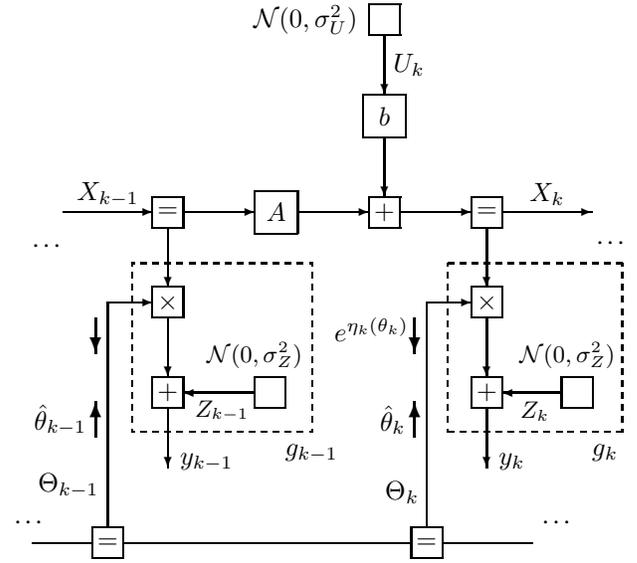
\begin{figure}
\begin{center}
\setlength{\unitlength}{0.8mm}
\begin{picture}(93,92)(0,0)
%
\put(0,52){\pos{l}{\ldots}}
\put(-3,6){\pos{l}{\ldots}}
\put(5,57.5){\vector(1,0){15}}        \put(12.5,60.5){\cent{$X_{k-1}$}}
\put(20,55){\framebox(5,5){$=$}}
\put(22.5,55){\vector(0,-1){10}}
\put(25,57.5){\vector(1,0){12}}
\put(37,54){\framebox(7,7){$A$}}
\put(44,57.5){\vector(1,0){12}}
\put(56,87){\framebox(5,5){}}         \put(54,89.5){\pos{r}{$\calN(0,\sigma_U^2)$}}
\put(58.5,87){\vector(0,-1){10}}      \put(60,82){\pos{l}{$U_k$}}
\put(55,70){\framebox(7,7){$b$}}
\put(58.5,70){\vector(0,-1){10}}
\put(56,55){\framebox(5,5){$+$}}
\put(61,57.5){\vector(1,0){12}}
\put(73,55){\framebox(5,5){$=$}}
\put(75.5,55){\vector(0,-1){10}}
\put(78,57.5){\vector(1,0){15}}       \put(85.5,60.5){\cent{$X_k$}}
\put(99,52.5){\pos{r}{\ldots}}
\put(90,6){\pos{r}{\ldots}}

\put(16.5,21){\dashbox(30,28){}}      \put(42,17){$g_{k-1}$}
\put(12.5,42.5){\vector(1,0){7.5}}
\put(20,40){\framebox(5,5){$\times$}}
\put(22.5,40){\vector(0,-1){10}}
\put(20,25){\framebox(5,5){$+$}}
 \put(37,25){\framebox(5,5){}}        
                                      \put(29,32.5){\small$\calN(0,\sigma_Z^2)$}
 \put(37,27.5){\vector(-1,0){12}}     \put(31.5,24.5){\cent{\small$Z_{k-1}$}}
\put(22.5,25){\vector(0,-1){10}}      \put(24.5,16){\pos{l}{$y_{k-1}$}} 
\put(0,2.5){\line(1,0){10}}
\put(12.5,5){\line(0,1){37.5}}        \put(11,12){\pos{r}{$\Theta_{k-1}$}}
 {\thicklines
  \put(10.5,40){\vector(0,-1){6}}     
  \put(10.5,20){\vector(0,1){6}}      \put(9,23){\pos{r}{$\hat{\theta}_{k-1}$}}
 }
\put(10,0){\framebox(5,5){$=$}}
\put(15,2.5){\line(1,0){48}}          

\put(69,21){\dashbox(29,28){}}        \put(93,17){$g_k$}
\put(65.5,42.5){\vector(1,0){7.5}}
\put(73,40){\framebox(5,5){$\times$}}
\put(75.5,40){\vector(0,-1){10}}
\put(73,25){\framebox(5,5){$+$}}
 \put(88,25){\framebox(5,5){}}        
                                      \put(81,32.5){\small $\calN(0,\sigma_Z^2)$}
 \put(88,27.5){\vector(-1,0){10}}     \put(83.5,24.5){\cent{\small$Z_k$}}
\put(75.5,25){\vector(0,-1){10}}      \put(77.5,16){\pos{l}{$y_{k}$}} 
\put(65.5,5){\line(0,1){37.5}}        \put(64.2,11){\pos{r}{$\Theta_{k}$}}
 {\thicklines
  \put(63.5,40){\vector(0,-1){6}}     \put(62.5,38){\pos{r}{$e^{\eta_k(\theta_k)}$}}
  \put(63.5,20){\vector(0,1){6}}      \put(62,23){\pos{r}{$\hat{\theta}_k$}}
 }
\put(63,0){\framebox(5,5){$=$}}
\put(68,2.5){\line(1,0){15}}
\end{picture}
\vspace{2mm}
\caption{\label{fig:StateSpaceModelWithUnknownCoefficient}%
Linear state space model with unknown coefficient vector 
$\Theta =\Theta_1=\Theta_2=\ldots$
and white Gaussian input signal $U_1,U_2,\ldots$
The figure shows one section of the factor graph.
The multiplier node denotes the inner product $\Theta_k^T X_k$.
The label $\calN(m,\sigma^2)$ denotes a scalar Gaussian factor with mean $m$ and variance $\sigma^2$.
The EM message computation rule is applied to the dashed boxes.}
\end{center}
\end{figure}

The factor graph of this system model, 
i.e., of
\iftwocolumn{
 \begin{IEEEeqnarray}{rCl}
  \IEEEeqnarraymulticol{3}{l}{
  p(u,x,y,z \cond \theta)
 }\nonumber\\
}%
{
 \begin{IEEEeqnarray}{rCl}
 p(u,x,y,z \cond \theta)
}
 & = & p(x_0) \prod_{k=1}^N p(y_k \cond x_k, z_k, \theta) p(z_k) p(x_k \cond x_{k-1}, u_k) p(u_k),
       \IEEEeqnarraynumspace
\end{IEEEeqnarray}
is shown in \Fig{fig:StateSpaceModelWithUnknownCoefficient}. 
Note that the unknown coefficient vector $\Theta$ appears in copies 
$\Theta_k$, $k=1,2,\ldots, N$ (one copy for each time~$k$) 
with an equality constraint 
$\Theta_1=\ldots =\Theta_N$. 
Note also that the factors  
$p(x_k \cond x_{k-1}, u_k)$ and $p(y_k \cond x_k, z_k, \theta)$
express the constraints (\ref{eqn:NextStateFIR}) and (\ref{eqn:NoisyOutputFIR}), respectively;
only the scalar Gaussian factors 
$p(u_k)$ and $p(z_k)$
are ``soft'' factors without Dirac deltas.
The factor $p(x_0)$ (not shown in \Fig{fig:StateSpaceModelWithUnknownCoefficient}) 
is of secondary importance and may even be omitted in practice.

Note that the edges $\Theta_k$, $k=1,2,\ldots,$
cut the factor graph into two cycle-free components. 
The equality constraints $\Theta_1=\Theta_2=\ldots$ at the bottom 
of \Fig{fig:StateSpaceModelWithUnknownCoefficient} 
correspond to $\fA$ in Figures \ref{fig:TwoFactorsEM} and~\ref{fig:GenStateSpaceEM};
everything else in \Fig{fig:StateSpaceModelWithUnknownCoefficient}
corresponds to $\fB$ in Figures \ref{fig:TwoFactorsEM} and~\ref{fig:GenStateSpaceEM}.

With estimates $\hat{\theta}_k$ plugged in, the upper part (the $\fB$ part)
of \Fig{fig:StateSpaceModelWithUnknownCoefficient} becomes 
a standard linear Gaussian factor graph, where sum-product message passing 
amounts to Kalman filtering\andor{}smoothing \cite[Section~V]{LDHKLK:fgsp2007}.

We now need to compute the EM messages $e^{\eta_k(\theta_k)}$. 
Heeding the advice of Section~\ref{sec:IssueHardConstraintsGrouping},
we group the multiplier node (which is a hard constraint) 
with the adjacent soft node\andor{}factor 
$p(z_k) \propto e^{-z^2/(2\sigma_Z^2)}$
as indicated by the dashed boxes in \Fig{fig:StateSpaceModelWithUnknownCoefficient};
this grouping 
(and integrating\andor{}marginalizing over the variables inside the box) 
results in the factor 
\iftwocolumn{
 \begin{IEEEeqnarray}{rCl}
 \IEEEeqnarraymulticol{3}{l}{
 g_k(x_k,y_k,\theta_k)
 }\nonumber\\\quad
}%
{\begin{IEEEeqnarray}{rCl}
 g_k(x_k,y_k,\theta_k)
}
 & = & \int_{z_k} \delta(\theta_k^T x_k + z_k - y_k)\, 
       \frac{1}{\sqrt{2\pi} \sigma_Z} e^{-z_k^2/(2\sigma_Z^2)}\, dz_k
       \IEEEeqnarraynumspace \\
 & \propto & e^{-(\theta_k^Tx_k - y_k)^2 / (2\sigma_Z^2)},
             \label{eqn:ExFIRg}
\end{IEEEeqnarray}
which is perfectly 
well-behaved.
Note that the missing scale factor in (\ref{eqn:ExFIRg}) 
can be safely ignored, cf.\ Table~\ref{tab:ElogMessageGeneral}.

As it turns out, the EM message $e^{\eta_k(\theta_k)}$ 
out of the dashed box $g_k$ 
in \Fig{fig:StateSpaceModelWithUnknownCoefficient}
is Gaussian 
with weight matrix (inverse covariance matrix) 
$\msgb{W}{\Theta_k}$ and mean vector $\msgb{m}{\Theta_k}$ 
as given by (\ref{eqn:GaussianEMFIRmsgbW})-(\ref{eqn:GaussianEMFIRWXmX}) 
in Table~\ref{tab:InnerProductBackwardsGaussianEM}
with $m_S=y_k$ and $\sigma_S^2=\sigma_Z^2$.
The proof of (\ref{eqn:GaussianEMFIRmsgbW})-(\ref{eqn:GaussianEMFIRWXmX})
is given in Section~\ref{sec:GaussianMultiplierEM}.

\begin{eqntable}
\caption{\label{tab:InnerProductBackwardsGaussianEM}%
Gaussian message passing backwards through a multiplier. 
$X$~and $\Theta$ are real column vectors 
and $S=\Theta^T X$ is a scalar. 
$\calN(m,\sigma^2)$ denotes a scalar Gaussian factor with mean $m$ and variance $\sigma^2$.
The incoming sum-product message $\msgf{\mu}{X}$ is Gaussian 
with parameters $\msgf{W}{X}$ and $\msgf{m}{X}$.
}
\begin{center}
\tablebox{
\begin{center}
\vspace{6mm}
\begin{picture}(47.5,25)(0,-2)
%
\put(12.5,-2){\dashbox(35,17){}}        
\put(0,7.5){\vector(1,0){17.5}}         \put(5,10){\cent{$X$}}
\put(17.5,5){\framebox(5,5){$\times$}}
\put(20,25){\vector(0,-1){15}}
{\thicklines
\put(18,22.5){\vector(0,-1){4.5}}      \put(16.5,20.5){\pos{r}{$\hat\theta$}}
\put(22,18){\vector(0,1){4.5}}         
									  \put(23.5,20.5){\pos{l}{$e^{\eta(\theta)}$}}
}
\put(22.5,7.5){\vector(1,0){15}}        \put(30,10){\cent{$S$}}
\put(37.5,5){\framebox(5,5){}}          
                                       \put(38,2){\cent{$\calN(m_S,\sigma_S^2)$}}
\end{picture}
\vspace{5mm}
\end{center}
$e^{\eta(\theta)}$ is Gaussian with
\begin{IEEEeqnarray}{rCl}
  \msgb{W}{\Theta} & = & \frac{V_X + m_Xm_X^T}{\sigma_S^2} 
                         \label{eqn:GaussianEMFIRmsgbW} \\
  \msgb{W}{\Theta} \msgb{m}{\Theta} &=& \frac{m_X m_S}{\sigma_S^2}
                         \label{eqn:GaussianEMFIRmsgbWm}
\end{IEEEeqnarray}
with $V_X$ and $m_X$ given by
\begin{IEEEeqnarray}{rCl}
  V_X^{-1} &=& \msgf{W}{X} + \hat{\theta}\, \hat{\theta}^T / \sigma_S^2 
               \label{eqn:GaussianEMFIRVX} \\
  W_X m_X  &=& \msgf{W}{X} \msgf{m}{X} + \hat{\theta}\, m_S / \sigma_S^2.
               \label{eqn:GaussianEMFIRWXmX}
\end{IEEEeqnarray}
\vspace{-2mm}
}
\end{center}
\end{eqntable}

It remains only to compute new estimates $\hat{\theta}_k$ 
by max-product message passing through the chain of equality constraints 
at the bottom of \Fig{fig:StateSpaceModelWithUnknownCoefficient}. 
Since the incoming EM messages $e^{\eta_k(\theta_k)}$ are Gaussians,
max-product message passing coincides with sum-product message passing 
with message computation rules as in Table~2 of~\cite{LDHKLK:fgsp2007}.

In summary, both the expectation step and the maximization step 
of the EM algorithm 
can be carried out by Gaussian message passing.

\subsection{Autoregressive Filter Identification}
\label{sec:AutoregressiveFilter}

Consider the following state space representation of an autoregressive model. 
Let the state $X_k\in\R^n$, $k=1,2,\ldots,N$ evolve according to
\begin{equation}
X_k = A X_{k-1} + b U_k
\end{equation}
with
\begin{equation}
b = (1,0,\ldots,0)^T
\end{equation}
and with $n\times n$ matrix
\begin{equation}
A(\Theta) = \left( \begin{array}{cc}
                   \multicolumn{2}{c}{\Theta^T} \\
                   I_{n-1} & 0
                  \end{array} \right)
               \label{eqn:AutregMatrix}
            \IEEEeqnarraynumspace
\end{equation}
where $\Theta$ is an unknown column vector of dimension~$n$.
We assume that the input signal $U_1, U_2,\ldots,$
which is often called ``innovation'',
is zero-mean white Gaussian noise with variance $\sigma_U^2$. 
We observe a noisy scalar output signal
\begin{equation}
Y_k = (1,0,\ldots,0)^T X_k + Z_k,
\end{equation}
where $Z_1,Z_2,\ldots$ is zero-mean white Gaussian noise with variance $\sigma_Z^2$.
From the observation $Y_k=y_k$, $k=1,2,\ldots,N$, we wish to estimate $\Theta$;
specifically, we wish to compute the maximum likelihood estimate
\begin{IEEEeqnarray}{rCl}
\hat{\theta}  & = &  \argmax_{\theta} p(y \cond \theta)
       \IEEEeqnarraynumspace\\
 & = & \argmax_{\theta} \int_u \int_x \int_z p(u,x,y,z \cond \theta)\, dz\, dx\, du
        \IEEEeqnarraynumspace
\end{IEEEeqnarray}
with $y\eqdef (y_1,y_2,\ldots,y_N)$ etc.

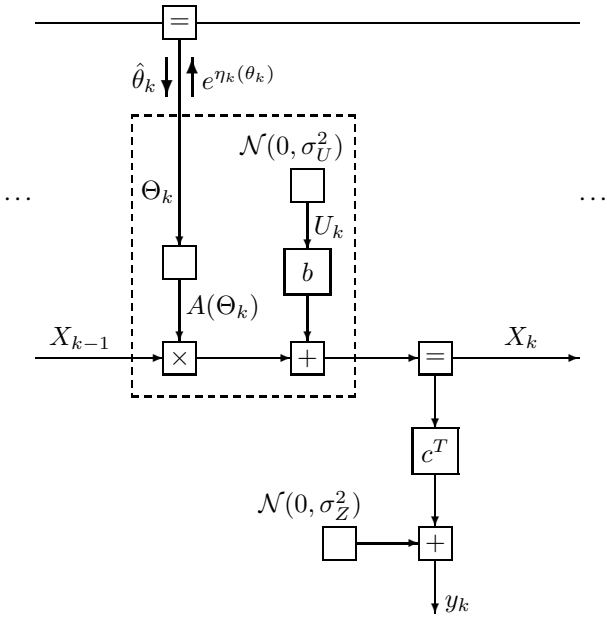
\begin{figure}
\begin{center}
\setlength{\unitlength}{0.85mm}
\begin{picture}(85,95)(0,0)
%
\put(0,65){\pos{r}{\ldots}}
\put(85,65){\pos{l}{\ldots}}
\put(0,92.5){\line(1,0){20}}
\put(20,90){\framebox(5,5){$=$}}
\put(25,92.5){\line(1,0){60}}
\put(22.5,90){\vector(0,-1){32.5}}      \put(21.7,66){\pos{r}{$\Theta_k$}}
 {\thicklines
  \put(20.5,87){\vector(0,-1){6}}       \put(19,84){\pos{r}{$\hat\theta_k$}}
  \put(24.5,81){\vector(0,1){6}}        \put(26,84){\pos{l}{$e^{\eta_k(\theta_k)}$}}
 }
\put(20,52.5){\framebox(5,5){}}
\put(22.5,52.5){\vector(0,-1){10}}      \put(23.5,47){$A(\Theta_k)$}
\put(15,34){\dashbox(35,44){}}
\put(0,40){\vector(1,0){20}}            \put(7,43){\cent{$X_{k-1}$}}
\put(20,37.5){\framebox(5,5){$\times$}}
\put(25,40){\vector(1,0){15}}
\put(40,37.5){\framebox(5,5){$+$}}
 \put(40,64.5){\framebox(5,5){}}        \put(32,72){$\calN(0,\sigma_U^2)$}
 \put(42.5,64.5){\vector(0,-1){7.5}}    \put(43.5,59.5){$U_k$}
 \put(39,50){\framebox(7,7){$b$}}
 \put(42.5,50){\vector(0,-1){7.5}}
\put(45,40){\vector(1,0){15}}
\put(60,37.5){\framebox(5,5){$=$}}
\put(65,40){\vector(1,0){20}}           \put(76,43.25){\cent{$X_k$}}
\put(62.5,37.5){\vector(0,-1){8.5}}
\put(59,22){\framebox(7,7){$c^T$}}
\put(62.5,22){\vector(0,-1){8.5}}
\put(60,8.5){\framebox(5,5){$+$}}
 \put(45,8.5){\framebox(5,5){}}         \put(43,17){\cent{$\calN(0,\sigma_{Z}^2)$}}
 \put(50,11){\vector(1,0){10}}          
\put(62.5,8.5){\vector(0,-1){8.5}}      \put(64,1){$y_k$}
\end{picture}
\vspace{2mm}
\caption{\label{fig:Autoregressive}%
Linear state space model for autoregressive filter 
with $b=c=(1,0,\ldots,0)^T$, 
with unknown coefficient vector $\Theta$, 
and with scalar white Gaussian innovation $U_1,U_2,\ldots$
The figure shows one section of the factor graph.
The multiplier node denotes the product $A(\Theta) X_k$ (\ref{eqn:AutregMatrix}). 
The EM message computation rule 
(\ref{eqn:TableAutregW}) and~(\ref{eqn:TableAutregWm}) 
applies to the dashed box.}
\end{center}
\end{figure}

The factor graph of $p(u,x,y,z \cond \theta)$ is shown in \Fig{fig:Autoregressive}. 
As in the previous example, 
the unknown parameter vector $\Theta$ appears in copies $\Theta_1 = \ldots = \Theta_N$,
one copy for each time~$k$.

Again, for fixed $\Theta=\hat{\theta}$, 
this factor graph is linear Gaussian and cycle-free. 

The EM message computation rule 
of Table~\ref{tab:ElogMessageGeneral} 
may be applied to the dashed box in \Fig{fig:Autoregressive}. 
It turns out that the EM message $e^{\eta_k(\theta_k)}$ 
is Gaussian with mean $\msgb{m}{\Theta_k}$ 
and weight matrix (inverse covariance matrix) $\msgb{W}{\Theta_k}$ 
given by (\ref{eqn:TableAutregW}) and (\ref{eqn:TableAutregWm}) in Table~\ref{tab:GaussMultEM}.

Again, we have obtained a purely Gaussian message passing algorithm. 
Apart from the EM message $e^{\eta_k(\theta_k)}$, 
all messages can be computed as described in~\cite[Section~V]{LDHKLK:fgsp2007}.

\subsection{Remarks}

We conclude this section with some remarks on these examples.
\begin{enumerate}
\item
In order to make the described algorithms work in practice, 
it is necessary to pay attention to the scheduling of the message updates. 
A serial (left-to-right) schedule may actually work better than alternating 
forward-backward sweeps in the two components 
(corresponding to $\fA$ and $\fB$) of the factor graph, 
cf.~\cite{Korl:diss2005}.
\item
The point of these examples is only to illustrate the message passing view 
of the EM algorithm; we are not concerned here with analyzing and 
comparing different approaches to 
linear-system identification \cite{Lj:sy}.
%
\item
Tabulated message computation rules 
(as in Table~\ref{tab:InnerProductBackwardsGaussianEM}) 
can greatly simplify the derivation 
of EM message passing algorithms.
\end{enumerate}

\section{Gaussian Message Passing\protect\iftwocolumn{\\ }{} Through Multiplier Nodes}
\label{sec:GaussianMultiplierEM}

A substantial part of traditional signal processing 
is essentially equivalent to 
Gaussian message passing in linear models \cite{LDHKLK:fgsp2007}. 
Unknown coefficients in such models 
introduce multiplier nodes into the corresponding factor graphs 
as is exemplified by Figures \ref{fig:StateSpaceModelWithUnknownCoefficient}
and~\ref{fig:Autoregressive}.

The EM message out of such multiplier nodes, 
properly grouped with ``soft'' Gaussian nodes\andor{}factors 
as in Figures \ref{fig:StateSpaceModelWithUnknownCoefficient}
and~\ref{fig:Autoregressive}, 
is invariably Gaussian 
(up to a scale factor), 
but the computation of its mean and its covariance matrix 
(in terms of the parameters of the incoming Gaussian messages) 
can be involved, 
cf.\ Appendices \ref{appsec:ProofsTableGaussMultEM}--\ref{appsec:VXY}. 
It is therefore helpful to tabulate such messages 
as exemplified by Table~\ref{tab:InnerProductBackwardsGaussianEM}.

However, such multiplier nodes come in surprisingly many versions:
scalar times scalar,
scalar times vector, 
inner product of two vectors (as in \Fig{fig:StateSpaceModelWithUnknownCoefficient}),
general matrix times vector, 
products involving matrices with a special structure (as in \Fig{fig:Autoregressive}), etc. 
Moreover, the grouping of such multiplier nodes with suitable soft factors\andor{}nodes 
is another source of virtually endless variety. 

We will therefore confine ourselves to a small number of cases which 
appear to be particulary useful and widely applicable. 
The general setup 
is shown in Table~\ref{tab:GaussMultEM}
and the results are given in Tables \ref{tab:GaussMultEM} and~\ref{tab:VXYetc}.
In all cases, we have a multiplier 
$U=A(\Theta) X$, where $A(\Theta)$ is a matrix that depends on~$\Theta$, 
grouped with $Y=U+Z$, where $Z$ is zero-mean Gaussian
with covariance matrix $V_Z=W_Z^{-1}$ (or $\sigma_Z^2$ in the scalar case).
In all cases, we assume that Gaussian messages $\msgf{\mu}{X}$ and $\msgb{\mu}{Y}$ 
arrive via the edges $X$ and $Y$, respectively;
these incoming messages are parameterized by
the mean vectors $\msgf{m}{X}$ and $\msgb{m}{Y}$
and the covariance matrices $\msgf{V}{X}=\msgf{W}{X}^{-1}$ and $\msgb{V}{Y}=\msgb{W}{Y}^{-1}$,
respectively.
The following cases are considered:
\begin{enumerate}
\item
Inner product:
$A(\Theta)=\Theta^T$, 
both $\Theta$ and $X$ are real column vectors (of the same dimension), 
and both $U=\Theta^T X$ and $Y$ are real scalars.

This case is a generalization of Table~\ref{tab:InnerProductBackwardsGaussianEM},
as will be discussed at the end of this section.
\item
Real scalar $\Theta$ times real column vector $X$: 
$A(\Theta)=\Theta$ and 
both $U=\Theta X$ and $Y$ are column vectors.

Some pertinent properties of the trace operator (``$\tr$'') 
are recalled in Appendix~\ref{appsec:Trace}.
\item
Componentwise product (denoted by $\Theta \odot X$) 
of real column vectors $\Theta$ and~$X$:
$A(\Theta)=\diag(\Theta)$, a diagonal matrix with the elements of $\Theta$ on the diagonal, 
and both $U=\Theta\odot X$ and $Y$ are column vectors.
\item
Autoregression: 
$\Theta, X, Y$ are column vectors in $\R^n$ and 
$A(\Theta)$ is the square matrix 
(\ref{eqn:AutregMatrix}) (which is essentially a companion matrix).
In addition, $Z$ is a zero-mean Gaussian vector with covariance matrix
\begin{equation} \label{eqn:AutregVZ}
V_Z = \left(\begin{array}{cccc} 
              \sigma_Z^2 & 0 & \ldots & 0 \\
                  0      & 0 &        & \vdots \\
               \vdots    &   & \ddots
            \end{array}\right),
\end{equation}
i.e., $Z$ is effectively a scalar that affects only the first component $Y_1$ of $Y$.
\item
General real matrix $\Theta$ times real column vector $X$: 
$A(\Theta)=\Theta$ and 
both $U=\Theta X$ and $Y$ are column vectors.

The symbol ``$\otimes$'' 
in (\ref{eqn:TableGenMatrixTimesVectW}) and~(\ref{eqn:TableGenMatrixTimesVectWm})
denotes the Kronecker product, 
cf.~(\ref{eqn:BeforeKronecker})--(\ref{eqn:QuadraticFormOfMatrixRvectW}).
More about this case is said below.
\end{enumerate}
The case of scalar $\Theta$ times scalar $X$ is a common special case
of all these cases
and does not need to be considered separately.

\begin{eqntable}
\caption{\label{tab:GaussMultEM}%
Gaussian backward EM~messages $e^{\eta(\theta)}$ through some multiplier nodes, 
see Section~\ref{sec:GaussianMultiplierEM}.
The EM message $e^{\eta(\theta)}$ is always Gaussian 
(up to a constant scale factor)
with parameters $\msgb{W}{\Theta}$ and $\msgb{m}{\Theta}$ as stated. 
See also Table~\ref{tab:VXYetc}.}
\begin{center}
\tablebox{
\begin{center}
\setlength{\unitlength}{0.9mm}
\vspace{6mm}
\begin{picture}(65,43)(0,0)
%
\put(0,7.5){\vector(1,0){20}}         \put(6,10.5){\cent{$X$}}
\put(22.5,43){\vector(0,-1){18}}      \put(24,38.5){\pos{l}{$\Theta$}}
\put(22.5,20){\vector(0,-1){10}}      \put(23.5,15){\pos{l}{$A(\Theta)$}}
\put(20,20){\framebox(5,5){}}
\put(20,5){\framebox(5,5){$\times$}}
\put(25,7.5){\vector(1,0){15}}        \put(32.5,10.5){\cent{$U$}}
 \put(40,20){\framebox(5,5){}}        \put(39,28.5){\cent{$\calN(0,V_Z)$}}
 \put(42.5,20){\vector(0,-1){10}}     \put(44,15){\pos{l}{$Z$}}
\put(40,5){\framebox(5,5){$+$}}
\put(45,7.5){\vector(1,0){20}}        \put(59,10.5){\cent{$Y$}}
\put(15,0){\dashbox(35,33){}}         \put(32.5,-4){\cent{$g(x,y,\theta)$}}
\end{picture}
\vspace{10mm}
\end{center}
Inner product $\Theta^T X$ of column vectors $\Theta$ and~$X$,\\
$A(\Theta)=\Theta^T$:
\begin{IEEEeqnarray}{rCl}
\msgb{W}{\Theta} & = & \sigma_Z^{-2} (V_X + m_X m_X^T) 
          \label{eqn:GaussianInnerProductW} \\
\msgb{W}{\Theta} \msgb{m}{\Theta} & = & \sigma_Z^{-2} (V_{XY} + m_X m_Y)
          \label{eqn:GaussianInnerProductWm}
\end{IEEEeqnarray}
\vspace{0mm}

\noindent%
Scalar $\Theta$ times column vector~$X$, $A(\Theta)=\Theta$:
\begin{IEEEeqnarray}{rCl}
1/\msgb{\sigma}{\Theta}^{2}  
  & = &  \tr\left( W_Z V_X \right) + m_X^T W_Z m_X
         \label{eqn:EMMultScalarTimesVectorWTrace}\\
\msgb{m}{\Theta} / \msgb{\sigma}{\Theta}^{2}
  & = &  \tr\left( W_Z V_{XY^T} \right) + m_X^T W_Z m_Y
         \label{eqn:EMMultScalarTimesVectorWmTrace}
\end{IEEEeqnarray}
\vspace{0mm}

\noindent%
Componentwise product $\Theta \odot X$ of column vectors $\Theta$ and~$X$,
$A(\Theta)=\diag(\Theta)$:
\begin{IEEEeqnarray}{rCl}
\msgb{W}{\Theta} & = &  W_Z \odot \left( V_X + m_X m_X^T \right) \\
\msgb{W}{\Theta} \msgb{m}{\Theta} 
 & = & \left( W_Z \odot \left( V_{XY^T} + m_X m_Y^T \right) \right) 
        \nonumber\IEEEeqnarraynumspace\\
        && \cdot \left(1,1,\ldots,1\right)^T
\end{IEEEeqnarray}
\vspace{0mm}

\noindent%
Autoregression, see (\ref{eqn:AutregMatrix}) and~(\ref{eqn:AutregVZ}):
\begin{IEEEeqnarray}{rCl}
\msgb{W}{\Theta} & = &  \sigma_Z^{-2} \left( V_X + m_X m_X^T \right) 
       \label{eqn:TableAutregW}\\
\msgb{W}{\Theta} \msgb{m}{\Theta}  & = & \sigma_Z^{-2} \left( V_{XY_1} + m_X m_{Y_1} \right)
       \label{eqn:TableAutregWm}
\end{IEEEeqnarray}
\vspace{0mm}

\noindent%
General matrix $\Theta$ times column vector~$X$, $A(\Theta)=\Theta$:\\
$e^{\eta(\theta)}$ is Gaussian in $\rvect(\theta)^T$ with
\begin{IEEEeqnarray}{rCl}
\msgb{W}{\Theta} & = &  W_Z \otimes (V_X + m_X m_X^T) 
        \label{eqn:TableGenMatrixTimesVectW}\\
\msgb{W}{\Theta} \msgb{m}{\Theta}
 & = & (W_Z \otimes I_n) \cvect(V_{XY^T} + m_X m_Y^T)
        \label{eqn:TableGenMatrixTimesVectWm}
       \IEEEeqnarraynumspace
\end{IEEEeqnarray}
\vspace{-2mm}
}
\end{center}
\end{eqntable}

In the cases 1--4, where $\Theta$ is a column vector (or a scalar), 
the EM message $e^{\eta(\theta)}$
is Gaussian with mean vector $\msgb{m}{\Theta}$ 
and weight matrix (inverse covariance matrix) $\msgb{W}{\Theta}$
as given in Table~\ref{tab:GaussMultEM}. 

In Case~5, where $\Theta$ is a matrix, 
we need the following notation. 
Let $B$ be any $m\times n$ matrix and let
\begin{equation}
B = \left(\begin{array}{c} b_1 \\ \vdots\\ b_m \end{array}\right)
\end{equation}
be the decomposition of $B$ into its rows.
We will use both the row stack vector
\begin{equation} \label{eqn:DefRowStack}
\rvect(B) \eqdef \left( b_1, \ldots, b_m \right) 
\end{equation}
and the analogous column stack vector $\cvect(B)$,
where the columns of $B$ are stacked into one column vector. 
For example, if 
\begin{equation}
B = \left( \begin{array}{cc} b_{1,1} & b_{1,2} \\ b_{2,1} & b_{2,2} \end{array} \right)
\end{equation}
then
$\rvect(B) = \left( b_{1,1}, b_{1,2}, b_{2,1}, b_{2,2} \right)$
and
$\cvect(B) = \left( b_{1,1}, b_{2,1}, b_{1,2}, b_{2,2} \right)^T$.
With this notation, the EM message is Gaussian in $\rvect(\Theta)^T$
with parameters (\ref{eqn:TableGenMatrixTimesVectW}) and~(\ref{eqn:TableGenMatrixTimesVectWm})
(see also (\ref{eqn:GenMatrixProofEta})). 

Note that Table~\ref{tab:GaussMultEM} gives the analog of 
(\ref{eqn:GaussianEMFIRmsgbW}) and~(\ref{eqn:GaussianEMFIRmsgbWm}) 
in Table~\ref{tab:InnerProductBackwardsGaussianEM}; 
the analog of (\ref{eqn:GaussianEMFIRVX}) and~(\ref{eqn:GaussianEMFIRWXmX}) 
is Table~\ref{tab:VXYetc}, 
which gives expressions for the marginal means $m_X$ and $m_Y$ 
and for the covariance matrices $V_{X}$ and $V_{XY^T}$ 
for fixed $\Theta=\hat{\theta}$
in terms of the parameters 
$\msgf{m}{X}$ and $\msgf{V}{X}$ 
and $\msgb{m}{Y}$ and $\msgb{V}{Y}$ 
of the incoming Gaussian sum-product messages. 
Note that Table~\ref{tab:VXYetc}
applies to all the cases in Table~\ref{tab:GaussMultEM} simultaneously.

The proofs of the claims in Table~\ref{tab:GaussMultEM} are given in 
Appendix~\ref{appsec:ProofsTableGaussMultEM} 
and the proofs of the claims in Table~\ref{tab:VXYetc} 
are given in appendices \ref{appsec:ProofsTableVxmx} and~\ref{appsec:VXY}.
Not surprisingly, some of these derivations are essentially equivalent 
to similar computations in the EM literature 
\cite{ShSt:TimeSeriesEM1982,RoGha:LinGaussRev1999,GiNi:rmldynsys2005}. 
Nevertheless, most of the statements in 
Tables~\ref{tab:GaussMultEM} and~\ref{tab:VXYetc} 
do not seem to be readily available in the prior literature.

We conclude this section by 
considering the specialization of Case~1 (inner product) 
to $Y=y$ fixed, which results in the situation of 
Table~\ref{tab:InnerProductBackwardsGaussianEM}. 
In this case, we have
\begin{equation}
m_Y = \msgb{m}{y} =y
\end{equation}
and
\begin{equation}
V_{XY} = V_{Y} = \msgb{V}{Y} = 0.
\end{equation}
With the translations $m_S=m_Y$ and $\sigma_S^2=\sigma_Z^2$,
it is obvious that (\ref{eqn:GaussianInnerProductW}) 
and (\ref{eqn:GaussianInnerProductWm}) 
specialize to (\ref{eqn:GaussianEMFIRmsgbW}) and (\ref{eqn:GaussianEMFIRmsgbWm}), respectively.
Moreover, with $A(\hat{\theta})^T = \hat{\theta}$, 
it is obvious that 
(\ref{eqn:GaussianEMFIRVX}) follows from (\ref{eqn:MultWX})
and
(\ref{eqn:GaussianEMFIRWXmX}) follows from (\ref{eqn:MultmX}).

\begin{eqntable}
\caption{\label{tab:VXYetc}%
Computation of means $m_X$ and $m_Y$ and covariance matrices $V_X$ and $V_{XY^T}$ 
in Table~\ref{tab:GaussMultEM}.}
\begin{center}
\tablebox{
\vspace{3mm}
%
Auxiliary quantities:
\begin{IEEEeqnarray}{rCl}
W_X 
 & = & \msgf{W}{X} + A(\hat{\theta})^T \! \left( V_Z + \msgb{V}{Y} \right)^{\!-1} \!\! A(\hat{\theta})
       \label{eqn:MultWX}
       \\[1ex]
\msgf{V}{Y}
 & = & A(\hat{\theta}) \msgf{V}{\!X} A(\hat{\theta})^T + V_Z 
       \IEEEeqnarraynumspace \label{eqn:msgfVy} 
       \\[1ex]
\tilde{W}_Y 
 & = & \left(\msgf{V}{Y} + \msgb{V}{Y}\right)^{-1} \label{eqn:tildeW}
\end{IEEEeqnarray}
Quantities in Table~\ref{tab:GaussMultEM}:
\begin{IEEEeqnarray}{rCl}
V_X
 & = & W_X^{-1} \\
 & = & \msgf{V}{X} - \msgf{V}{X} A(\hat{\theta})^T \tilde{W}_Y A(\hat{\theta}) \msgf{V}{X}
       \IEEEeqnarraynumspace \label{eqn:VX}
       \\[1ex]
V_{XY^T} 
 & = & \msgf{V}{\!X} A(\hat{\theta})^T \tilde{W}_Y \msgb{V}{Y}
       \IEEEeqnarraynumspace \label{eqn:VXY}
       \\[1ex]
m_X
 & = &  V_X \bigg( \msgf{W}{X} \msgf{m}{X} 
          + A(\hat{\theta})^T \left( V_Z + \msgb{V}{Y} \right)^{-1} \msgb{m}{Y} \bigg)
        \nonumber \\
        && \label{eqn:MultmX} \\
 & = &  \left( I_n - \msgf{V}{X} A(\hat{\theta})^T \tilde{W}_Y A(\hat{\theta}) \right)
        \nonumber\\ && \cdot{}
        \left( \msgf{m}{X} + \msgf{V}{X} A(\hat{\theta})^T \left( V_Z + \msgb{V}{Y} \right)^{-1} \msgb{m}{Y} \right)
        \nonumber \\
        \label{eqn:MultmXalt}
       \\[1ex]
m_Y
 & = & V_Y \left( \msgf{W}{Y} \msgf{m}{Y} + \msgb{W}{Y} \msgb{m}{Y} \right)
          \IEEEeqnarraynumspace \label{eqn:MultmY} \\
 & = &  \left( I_m - \msgf{V}{Y} \tilde{W}_Y \right)
           \left( \msgf{m}{Y} + \msgf{V}{Y} \msgb{W}{Y} \msgb{m}{Y} \right)
           \IEEEeqnarraynumspace \label{eqn:MultmYalt}
\end{IEEEeqnarray}
\vspace{-2mm}
}
\end{center}
\end{eqntable}

\section{Conclusions}
\label{sec:Conclusions}

We have showed that EM may be viewed 
and used as a message passing algorithm in factor graphs,
and we have identified a general ``local'' EM message computation rule 
(Table~\ref{tab:ElogMessageGeneral}). 
In some important cases, the EM messages are tractable expressions, 
which was exemplified by the EM message out of multipliers
(arising from unknown coefficients) 
in linear Gaussian models.

As a full member of the family of message passing algorithms, 
it is easy to seamlessly combine expectation maximization with 
other message passing algorithms in interesting ways. 
In particular:
\begin{itemize}
\item
EM messages (like all messages) may be represented in many different ways
(including 
Gaussians as in Sections~\ref{sec:Examples} and~\ref{sec:GaussianMultiplierEM}, 
Gaussian mixtures \cite{DEKL:EMId}, particles \cite{DKL:ISIT2006c}, etc.,
leading to quite different actual computations. 
\item
The freedom (or the necessity) to choose some definite 
message update schedule leads to different algorithms with different performance;
more about this will be said in~\cite{DEKL:EMId}. 
\item
The maximization step amounts to applying the max-product algorithm 
to the corresponding subgraph, which in turn may be carried out by many 
(exact or approxiate) message passing algorithms.
For example, in some important applications (as, e.g., in Section~\ref{sec:Examples}), 
the maximization step can be done by Kalman filtering\andor{}smoothing. 
\item
The expectation step relies on plain sum-product messages. 
However, depending on the involved nodes and message types, the sum-product 
algorithm may be realized (exactly or approximately) in many different ways, 
cf.\ \cite[Section~VI]{LDHKLK:fgsp2007}. 
\end{itemize}

Moreover,
it is a general observation that tabulated message computation rules 
can greatly simplify the derivation of message passing algorithms \cite{LDHKLK:fgsp2007}.
This applies, in particular, to EM messages, 
which we have tabulated for various multiplier nodes (scalar, vector, general matrix, \ldots)
with incoming Gaussian messages. 
With these message tables, EM algorithms for 
a number of basic linear-system identification problems 
can easily be composed without additional derivations or computations.
More such tables will be given in \cite{DEKL:EMId}.

%


\appendices



\section{Proof of Equation~(\ref{eqn:EMTheorem})}
\label{appsec:ProofEM}

We give a variation of a standard proof (cf.\ \cite{StSe:cmem2004}) 
that is adapted to the setup of Section~\ref{sec:EM}.
The heart of the proof is the following fact.

\begin{trivlist}\item[]{\bfseries Lemma:}
The function
\begin{equation} \label{eqn:EMLemmaAuxFunc}
\tilde f(\theta,\hat{\theta}) 
\eqdef f(\hat{\theta}) + \int_x f(x,\hat{\theta}) \log\!\left( \frac{f(x,\theta)}{f(x,\hat{\theta})}\right) \, dx
\end{equation}
(where ``$\log$'' denotes the natural logarithm)
satisfies both
\begin{equation} \label{eqn:EMLemmaAuxFuncIneq}
\tilde f(\theta,\hat{\theta}) \leq f(\theta)
\end{equation}
and
\begin{equation} \label{eqn:EMLemmaAuxFuncEq}
\tilde f(\theta,\theta) = f(\theta).
\end{equation}
\eproofnegspace\hfill$\Box$
\end{trivlist}
 
\begin{IEEEproof}
The equality (\ref{eqn:EMLemmaAuxFuncEq}) is obvious. 
The inequality (\ref{eqn:EMLemmaAuxFuncIneq}) follows from 
eliminating the logarithm in (\ref{eqn:EMLemmaAuxFunc}) by 
the inequality $\log(x) \leq x-1$ for $x>0$:
\begin{IEEEeqnarray}{rCl}
 \tilde f(\theta,\hat{\theta}) 
& \leq &  f(\hat{\theta}) + \int_x f(x,\hat{\theta}) 
             \left( \frac{f(x,\theta)}{f(x,\hat{\theta})} - 1 \right) dx 
          \IEEEeqnarraynumspace \\
& = &     f(\hat{\theta}) + \int_x f(x,\theta)\, dx - \int_x f(x,\hat{\theta})\, dx \\
& = &     f(\theta).
\end{IEEEeqnarray}
\eproofnegspace%
\end{IEEEproof}

\noindent
To prove~(\ref{eqn:EMTheorem}),
we first note that (\ref{eqn:EMmaxstep}) is equivalent to
\begin{equation} \label{eqn:maxstepAuxFunc}
\smpl{\hat\theta}{k+1} = \argmax_\theta \tilde f(\theta,\smpl{\hat\theta}{k}).
\end{equation}
We then obtain
\begin{IEEEeqnarray}{rCl}
f(\smpl{\hat\theta}{k}) 
& = &    \tilde f(\smpl{\hat\theta}{k}, \smpl{\hat\theta}{k})  \label{eqn:ProofEMLemmaI}\\
& \leq & \tilde f(\smpl{\hat\theta}{k+1}, \smpl{\hat\theta}{k}) \label{eqn:ProofEMLemmaII}\\
& \leq & f(\smpl{\hat\theta}{k+1}), \label{eqn:ProofEMLemmaIII}
\end{IEEEeqnarray}
where (\ref{eqn:ProofEMLemmaI}) follows from (\ref{eqn:EMLemmaAuxFuncEq}), 
(\ref{eqn:ProofEMLemmaII}) follows from (\ref{eqn:maxstepAuxFunc}), 
and (\ref{eqn:ProofEMLemmaIII}) follows from (\ref{eqn:EMLemmaAuxFuncIneq}).

\section{Some Properties of the Trace Operator}
\label{appsec:Trace}

We recall some pertinent properties of the trace operator 
for use in Appendix~\ref{appsec:ProofMultScalarTimesVector}.
The entries of a matrix $A$ will be denoted by $a_{k,\ell}$. 
The trace of a square matrix $A$ is the sum of the diagonal elements of $A$:
\begin{equation}
\tr(A) \eqdef \sum_k a_{k,k}.
\end{equation}
For matrices $A$ and $B$ such that $AB$ is a square matrix
(i.e., $B$ has the same dimensions as $A^T$), we have
\begin{IEEEeqnarray}{rCl}
\tr(AB) 
 & = &  \sum_k \sum_{\ell} a_{k,\ell} b_{\ell,k} 
        \IEEEeqnarraynumspace\\
 & = &  \tr(BA).  \label{eqn:TraceAB}
\end{IEEEeqnarray}
In particular, 
if $x$ and $y$ are column vectors (with the same number of rows), we have 
\begin{equation}
x^T y = y^T x = \tr(xy^T).
\end{equation}
Moreover, for $W=A^TA$, we have
\begin{IEEEeqnarray}{rCl}
x^T W y
 & = &  (Ax)^T Ay \\
 & = &  \tr(Ax (Ay)^T) \\
 & = &  \tr(A xy^T\! A^T),
        \IEEEeqnarraynumspace
\end{IEEEeqnarray}
and using (\ref{eqn:TraceAB}) we further obtain
\begin{IEEEeqnarray}{rCl}
x^T W y 
 & = & \tr(W xy^T) \\
 & = & \tr(xy^T W).
\end{IEEEeqnarray}
Now let $X$ and $Y$ be random column vectors with the same dimensions.
Let $m_X\eqdef \E[X]$ and $m_Y\eqdef \E[Y]$ and 
\begin{equation} \label{eqn:TraceVXYT}
V_{XY^T} \eqdef \EE{(X-m_X) (Y-m_Y)^T}.
\end{equation}
Then, for any square matrix $W$ as above (i.e., $W=A^T A$) with suitable dimensions, we have
\iftwocolumn{
 \begin{IEEEeqnarray}{rCl}
  \IEEEeqnarraymulticol{3}{l}{
  \EE{X^T W Y}
  }\nonumber\\\quad
}%
{
 \begin{IEEEeqnarray}{rCl}
 \EE{X^T W Y}
}
 & = &  \EE{(X-m_X)^T W (Y-m_Y)} + m_X^T W m_Y  \IEEEeqnarraynumspace\\
 & = &  \EE{\tr\!\left( W (X-m_X) (Y-m_Y)^T \right)}
         + m_X^T W m_Y \IEEEeqnarraynumspace\\
 & = &  \tr\left( W V_{XY^T} \right) + m_X^T W m_Y.
            \label{eqn:EXTWYTraceWVXYT}
\end{IEEEeqnarray}

\section{Proofs of the Claims in Table~\ref{tab:GaussMultEM}}
\label{appsec:ProofsTableGaussMultEM}

Recall (for repeated use below) that the probability density function
of an $n$-dimensional real Gaussian random vector
\begin{IEEEeqnarray}{rCl}
f(x) & = & \sqrt{\frac{\det(W)}{(2\pi)^n}}\,
           e^{-\frac{1}{2}(x-m)^T W (x-m)}
           \IEEEeqnarraynumspace \label{eqn:GenGaussian} \\
 & \propto & e^{-\frac{1}{2}(x^TWx-2x^TWm)},
             \IEEEeqnarraynumspace \label{eqn:GaussianxWxxWm}
\end{IEEEeqnarray}
where $m$ is the mean vector and $W=V^{-1}$ (a positive definite matrix) 
is the inverse of the covariance matrix~$V$.
In the scalar case ($n=1$), we will also use the notation $\sigma^2 \eqdef V$.


Now consider the factor graph in Table~\ref{tab:GaussMultEM}. 
The closed-box function $g(x,y,\theta)$
is obtained by marginalization\andor{}integration
over the variables inside the dashed box:
\iftwocolumn{
 \begin{IEEEeqnarray}{rCl}
  \IEEEeqnarraymulticol{3}{l}{
  g(x,y,\theta) 
  }\nonumber\\
}
{
 \begin{IEEEeqnarray}{rCl}
 g(x,y,\theta)
}
 & = & \int_u \delta(u-A(\theta) x) 
       \sqrt{\frac{\det(W_Z)}{(2\pi)^n}}\, e^{-\frac{1}{2}(y-u)^T W_Z (y-u)}\, du 
       \IEEEeqnarraynumspace\\
 & = & \sqrt{\frac{\det(W_Z)}{(2\pi)^n}}\, e^{-\frac{1}{2}(y-A(\theta) x)^T W_Z (y-A(\theta) x)}.
       \label{eqn:gxythetaProofs}
\end{IEEEeqnarray}
The exponent (\ref{eqn:GenLocalEMRuleExpect}) of the EM message $e^{\eta(\theta)}$ is 
\begin{IEEEeqnarray}{rCl}
\eta(\theta) 
 & = &  \EE{ \log g(X,Y,\theta) } \\
 & = &  \frac{1}{2} \log\!\left(\frac{\det(W_Z)}{(2\pi)^n}\right) 
        \iftwocolumn{\nonumber\\ && {}}{}
       - \frac{1}{2} \EE{(Y-A(\theta) X)^T W_Z (Y-A(\theta) X)} 
          \IEEEeqnarraynumspace \label{eqn:EMMultEtaGenExp2}\\
 & = &  \text{const}
           - \frac{1}{2} \Big( \EE{(A(\theta) X)^T W_Z (A(\theta) X)} 
          \iftwocolumn{\nonumber\\ && {~~~~~~~~~~~~~~~~~~~~~~~}}{}
        - 2\EE{(A(\theta) X)^T W_Z Y} \!\Big),
           \IEEEeqnarraynumspace \label{eqn:EMMultEtaGenExp}
\end{IEEEeqnarray}
where all logarithms are natural,
where the expectation is over $X$ and $Y$ 
(with respect to the local probability (\ref{eqn:plocalGenMult})), 
and where ``const'' subsumes all terms that do not depend on $\theta$. 

We are now ready to discuss the individual cases of Table~\ref{tab:GaussMultEM}.

\subsection{Inner Product $\Theta^T X$ of Column Vectors $\Theta$ and~$X$}

In this case, we have $A(\theta)=\theta^T$. 
The quantities $\theta^T X$, $Y$, and $W_Z$ are scalars;
in particular, $(\theta^T X)^T= \theta^TX$.
Thus (\ref{eqn:EMMultEtaGenExp}) becomes
\begin{IEEEeqnarray}{rCl}
\eta(\theta)
& = &  - \frac{1}{2} \Big( \EE{(\theta^T X)^T W_Z (\theta^T X)} 
                        - 2\EE{(\theta^T X)^T W_Z Y} \!\Big)
       \iftwocolumn{\nonumber\\ && {~~~~~~~~~~~~~~~~~~}}{}
       + \text{const} \IEEEeqnarraynumspace \\
& = &  - \frac{1}{2} \Big( \EE{\theta^T X W_Z X^T \theta} 
                        - 2\EE{\theta^T X W_Z Y} \!\Big)
       \iftwocolumn{\nonumber\\ && {~~~~~~~~~~~~~~~~~~}}{}
       + \text{const} \IEEEeqnarraynumspace \\
& = &  - \frac{1}{2} \Big( \theta^T \EE{X W_Z X^T} \theta
                        - 2\theta^T \EE{X W_Z Y} \!\Big)
       \iftwocolumn{\nonumber\\ && {~~~~~~~~~~~~~~~~~~}}{}
       + \text{const.} \IEEEeqnarraynumspace
\end{IEEEeqnarray}
It is then obvious from (\ref{eqn:GaussianxWxxWm})
that the EM message $e^{\eta(\theta)}$ is Gaussian 
(up to a scale factor) 
with weight matrix 
\begin{IEEEeqnarray}{rCl}
\msgb{W}{\Theta} 
 & = & \EE{XX^T}\sigma_Z^{-2} \\
 & = & \frac{V_X + m_X m_X^T}{\sigma_Z^2}
\end{IEEEeqnarray}
and
\begin{IEEEeqnarray}{rCl}
\msgb{W}{\Theta} \msgb{m}{\Theta} 
 & = & \EE{XY} \sigma_Z^{-2} \\
 & = & \frac{V_{XY} + m_X m_Y}{\sigma_Z^2}
\end{IEEEeqnarray}

\subsection{Scalar $\Theta$ Times Column Vector $X$}
\label{appsec:ProofMultScalarTimesVector}

In this case, we have $A(\theta)=\theta$, a scalar, 
and (\ref{eqn:EMMultEtaGenExp}) becomes 
\begin{equation}
\eta(\theta) = \text{const}
           - \frac{1}{2} \Big( \theta^2 \EE{X^T W_Z X} 
            - 2\theta\EE{X^T W_Z Y} \!\Big).
\end{equation}
It follows from (\ref{eqn:GaussianxWxxWm}) that $e^{\eta(\theta)}$ is Gaussian with 
\begin{IEEEeqnarray}{rCl}
\msgb{\sigma}{\Theta}^{-2} 
 & = &  \EE{X^TW_ZX} \\
 & = &  \tr\left( W_Z V_X \right) + m_X^T W_Z m_X
        \IEEEeqnarraynumspace
           \label{eqn:EMMultScalarTimesVectorWTraceProof}
\end{IEEEeqnarray}
and
\begin{IEEEeqnarray}{rCl}
\msgb{m}{\Theta} / \msgb{\sigma}{\Theta}^{2} 
 & = &  \EE{X^TW_Z Y} \\
 & = &  \tr\left( W_Z V_{XY^T} \right) + m_X^T W_Z m_Y
        \IEEEeqnarraynumspace
           \label{eqn:EMMultScalarTimesVectorWmTraceProof}
\end{IEEEeqnarray}
where (\ref{eqn:EMMultScalarTimesVectorWTraceProof}) 
and~(\ref{eqn:EMMultScalarTimesVectorWmTraceProof}) 
follow from (\ref{eqn:EXTWYTraceWVXYT})
and with $V_{XY^T}$ defined as in~(\ref{eqn:DefVXYT}).

\subsection{Componentwise Product $\Theta\odot X$ of Column Vectors}

In this case, we have $A(\theta) = \diag(\theta)$,
a diagonal matrix with the elements of $\theta$ on the diagonal,
and (\ref{eqn:EMMultEtaGenExp}) becomes 
\begin{IEEEeqnarray}{rCl}
\eta(\theta) 
 & = &  \text{const}
           - \frac{1}{2} \Big( \EE{(\diag(\theta) X)^T W_Z (\diag(\theta) X)} 
        \iftwocolumn{\nonumber\\ && {~~~~~~~~~~~~~~~~~~~~}}{}
       - 2\EE{(\diag(\theta) X)^T W_Z Y} \!\Big) 
           \IEEEeqnarraynumspace\\
 & = &  \text{const}
           - \frac{1}{2} \Big( \EE{(\diag(X) \theta)^T W_Z (\diag(X) \theta)} 
        \iftwocolumn{\nonumber\\ && {~~~~~~~~~~~~~~~~~~~~}}{}
       - 2\EE{(\diag(X) \theta)^T W_Z Y} \!\Big) 
           \IEEEeqnarraynumspace\\
 & = &  \text{const}
           - \frac{1}{2} \Big( \theta^T \EE{\diag(X) W_Z \diag(X)} \theta
        \iftwocolumn{\nonumber\\ && {~~~~~~~~~~~~~~~~~~~~}}{}
       - 2\theta^T \EE{\diag(X) W_Z Y} \!\Big).
          \IEEEeqnarraynumspace
\end{IEEEeqnarray}
It follows from (\ref{eqn:GaussianxWxxWm}) that $e^{\eta(\theta)}$ is Gaussian with
\begin{IEEEeqnarray}{rCl}
\msgb{W}{\Theta}
 & = &  \EE{\diag(X) W_Z \diag(X)} \IEEEeqnarraynumspace\\
 & = &  W_Z \odot \EE{X X^T} \IEEEeqnarraynumspace\\
 & = &  W_Z \odot \left( V_X + m_X m_X^T \right)
        \IEEEeqnarraynumspace
\end{IEEEeqnarray}
and
\begin{IEEEeqnarray}{rCl}
\msgb{W}{\Theta} \msgb{m}{\Theta} 
 & = &  \EE{\diag(X) W_Z Y}  \IEEEeqnarraynumspace\\
 & = &  \EE{\diag(X) W_Z \diag(Y)\left(1,1,\ldots,1\right)^T}  \IEEEeqnarraynumspace\\
 & = &  \left( W_Z \odot \EE{XY^T} \right)
           \left(1,1,\ldots,1\right)^T  \IEEEeqnarraynumspace\\
 & = &  \left( W_Z \odot \left( V_{XY^T} + m_X m_Y^T \right) \right)
           \left(1,1,\ldots,1\right)^T\!\! .
         \IEEEeqnarraynumspace
\end{IEEEeqnarray}

\subsection{Autoregression (Companion Matrix)}

In this case,
recall from (\ref{eqn:AutregMatrix}) that
\begin{equation}
A(\theta) \eqdef \left( \begin{array}{cc}
                   \multicolumn{2}{c}{\theta^T} \\
                   I_{n-1} & 0
                  \end{array} \right)
               \label{eqn:AutregMatrixProof}
\end{equation}
where $n$ is the dimension of the column vector $\theta$,
and where $I_{n-1}$ is the $(n-1) \times (n-1)$ identity matrix. 

Before we proceed, we need to address the following issue.
According to (\ref{eqn:AutregVZ}), we have
\begin{equation} \label{eqn:AutregVZEpsilon}
V_Z = \left( \begin{array}{ccccc}
        \sigma_Z^2 & 0 & 0 & \ldots & 0 \\
           0       & \varepsilon & 0 & \ldots & 0 \\
           0       & 0 & \varepsilon & \ldots \\
           \ldots\\
      \end{array} \right)
      \IEEEeqnarraynumspace
\end{equation}
with $\varepsilon = 0$, which creates a problem with $W_Z=V_Z^{-1}$.
We address this problem by proceeding with (\ref{eqn:AutregVZEpsilon}) 
with $\varepsilon>0$. As it turns out, the resulting 
expression for $\eta(\theta)$ does not depend on $\varepsilon$ 
(except in an additive constant, which we ignore). 

Using (\ref{eqn:AutregMatrixProof}), (\ref{eqn:EMMultEtaGenExp}) becomes
\begin{IEEEeqnarray}{rCl}
\eta(\theta) 
& = &  \text{const}
           - \frac{1}{2} \Bigg( \EE{\left(\begin{array}{c} \theta^TX \\ X_1 \\ \vdots \\ X_{n-1} \end{array}\right)^{\!\! T} 
              W_Z \left(\begin{array}{c} \theta^TX \\ X_1 \\ \vdots \\ X_{n-1} \end{array}\right) } 
       \iftwocolumn{\nonumber\\ && {~~~~~~~~~~~~~~~}}{}
        - 2\EE{\left(\begin{array}{c} \theta^TX \\ X_1 \\ \vdots \\ X_{n-1} \end{array}\right)^{\!\! T} \! W_Z Y} \!\Bigg).
           \IEEEeqnarraynumspace
\end{IEEEeqnarray}
Using (\ref{eqn:AutregVZEpsilon}) and ignoring all constant terms yields
\begin{IEEEeqnarray}{rCl}
\eta(\theta) 
& = &  \text{const} 
       \iftwocolumn{\nonumber\\ && {}}{}
       - \frac{1}{2} \Big( \EE{\theta^TX \sigma_Z^{-2} \theta^TX}
                 - 2\EE{\theta^TX \sigma_Z^{-2}Y_1} \Big)
       \IEEEeqnarraynumspace \\
& = &  \text{const} 
       \iftwocolumn{\nonumber\\ && {}}{}
       - \frac{1}{2} \Big( \theta^T \sigma_Z^{-2} \EE{XX^T} \theta
                 - 2\theta^T \sigma_Z^{-2} \EE{XY_1} \Big).
       \IEEEeqnarraynumspace
\end{IEEEeqnarray}
It follows from (\ref{eqn:GaussianxWxxWm}) that $e^{\eta(\theta)}$ is Gaussian with 
\begin{IEEEeqnarray}{rCl}
\msgb{W}{\Theta}
 & = &  \sigma_Z^{-2}\, \EE{X X^T} \\
 & = &  \sigma_Z^{-2} \left( V_X + m_X m_X^T \right)
        \IEEEeqnarraynumspace
\end{IEEEeqnarray}
and
\begin{IEEEeqnarray}{rCl}
\msgb{W}{\Theta} \msgb{m}{\Theta} 
 & = &  \sigma_Z^{-2}\, \EE{XY_1} \\
 & = &  \sigma_Z^{-2} \left( V_{XY_1} + m_X m_{Y_1} \right).
        \IEEEeqnarraynumspace
\end{IEEEeqnarray}

\subsection{General Matrix $\Theta$ Times Column Vector~$X$}
\label{Appsec:GenMatrix}

We need to begin with some preparations.
Recall the row stack operator $\rvect$ (\ref{eqn:DefRowStack}) and the corresponding
column stack operators $\cvect$.
Let $A$ be an $m\times n$ matrix with rows $a_1,\ldots, a_m$. 
For any column vector $x\in\R^n$ 
and any $m\times m$ square matrix~$W$ (with elements $w_{k,\ell}$), 
we have
\iftwocolumn{
 \begin{IEEEeqnarray}{rCl}
  \IEEEeqnarraymulticol{3}{l}{
  (Ax)^T W Ax
   = (a_1 x, \ldots, a_mx)\, W \!\left(\!\begin{array}{c} a_1x\\ \vdots\\ a_mx \end{array}\!\right)
  } \\\quad
}%
{
 \begin{IEEEeqnarray}{rCl}
 (Ax)^T W Ax
 & = & (a_1 x, \ldots, a_mx)\, W \!\left(\!\begin{array}{c} a_1x\\ \vdots\\ a_mx \end{array}\!\right) \\
}
 & = &  \sum_{k=1}^m \sum_{\ell=1}^m a_k x w_{k,\ell} (a_\ell x) \\
 & = &  \sum_{k=1}^m \sum_{\ell=1}^m a_k w_{k,\ell} x x^T\! a_\ell^T \\
 & = &  \left( a_1,\ldots, a_m \right)
        \iftwocolumn{\nonumber\\ && {}\cdot}{}
           \left(\begin{array}{ccc} 
                   w_{1,1}xx^T  & \ldots & w_{1,m}xx^T \\
                   \vdots & & \vdots \\
                   w_{m,1}xx^T  & \ldots & w_{m,m}xx^T
                 \end{array}\right)
           \left(\!\begin{array}{c} a_1^T\\ \vdots\\ a_m^T \end{array}\!\right) 
           \IEEEeqnarraynumspace \label{eqn:BeforeKronecker}\\
 & = &  \rvect(A) \left( W \otimes xx^T \right) \rvect(A)^T.
           \label{eqn:QuadraticFormOfMatrixRvectW}
\end{IEEEeqnarray}
Moreover, for any column vector $y\in\R^m$, 
we have
\iftwocolumn{
 \begin{IEEEeqnarray}{rCl}
  \IEEEeqnarraymulticol{3}{l}{
  (Ax)^T Wy
    = (a_1 x, \ldots, a_mx)\, W \!\left(\!\begin{array}{c} y_1\\ \vdots\\ y_m \end{array}\!\right)
  } \\\quad
}%
{
 \begin{IEEEeqnarray}{rCl}
  (Ax)^T Wy
   & = & (a_1 x, \ldots, a_mx)\, W \!\left(\!\begin{array}{c} y_1\\ \vdots\\ y_m \end{array}\!\right) \\
}
 & = &  \sum_{k=1}^m \sum_{\ell=1}^m a_k x w_{k,\ell} y_\ell \\
 & = &  \sum_{k=1}^m \sum_{\ell=1}^m a_k w_{k,\ell} x y_\ell \\
 & = &  \left( a_1,\ldots, a_m \right)
        \iftwocolumn{\nonumber\\ && {}\cdot}{}
           \left(\begin{array}{ccc} 
                   w_{1,1}I_n  & \ldots & w_{1,m}I_n \\
                   \vdots & & \vdots \\
                   w_{m,1}I_n  & \ldots & w_{m,m}I_n
                 \end{array}\right)
           \left(\!\begin{array}{c} xy_1\\ \vdots\\ xy_m \end{array}\!\right) 
           \IEEEeqnarraynumspace\\
 & = &   \rvect(A) \left( W \otimes I_n \right) \cvect(xy^T).
           \label{eqn:QuadraticFormOfMatrixRvectWXY}
\end{IEEEeqnarray}

After these preparations, we return to the EM message 
for the case where $A(\theta)=\Theta$ is a general $m\times n$ matrix. 
In this case, (\ref{eqn:EMMultEtaGenExp}) becomes
\begin{equation}
\eta(\Theta) = \text{const}
           - \frac{1}{2} \Big( \EE{(\Theta X)^T W_Z (\Theta X)} 
                               - 2\EE{(\Theta X)^T W_Z Y} \!\Big)
\end{equation}
and using (\ref{eqn:QuadraticFormOfMatrixRvectW}) and (\ref{eqn:QuadraticFormOfMatrixRvectWXY}) 
we obtain
\begin{IEEEeqnarray}{rCl}
\eta(\Theta)
 & = &  \text{const}
           - \frac{1}{2} \Big( \rvect(\Theta)\, \EE{W_Z \otimes XX^T} \rvect(\Theta)^T 
        \iftwocolumn{\nonumber\\ && {}}{}
        - 2\rvect(\Theta)\, \EE{(W_Z\otimes I_n) \cvect(XY^T)} \!\Big).
        \IEEEeqnarraynumspace \label{eqn:GenMatrixProofEta}
\end{IEEEeqnarray}
We now see that $e^{\eta(\Theta)}$ is Gaussian in $\rvect(\Theta)^T$
with
\begin{IEEEeqnarray}{rCl}
\msgb{W}{\Theta}
 & = &  W_Z \otimes \EE{XX^T}  \label{eqn:GenMatrixProofWa} \\
 & = &  W_Z \otimes (V_X + m_X m_X^T)
        \IEEEeqnarraynumspace
\end{IEEEeqnarray}
and
\begin{IEEEeqnarray}{rCl}
\msgb{W}{\Theta} \msgb{m}{\Theta}
 & = &  (W_Z \otimes I_n) \cvect(\EE{XY^T}) \\
 & = &  (W_Z \otimes I_n) \cvect(V_{XY^T} + m_X m_Y^T).
        \IEEEeqnarraynumspace \label{eqn:GenMatrixProofWm}
\end{IEEEeqnarray}

\section{Proofs of the Claims in Table~\ref{tab:VXYetc} Except~(\ref{eqn:VXY})}
\label{appsec:ProofsTableVxmx}

We consider the computation of
the mean vectors $m_X$ and $m_Y$ and the 
covariance matrix $V_{X}$
with respect to the local probability density (\ref{eqn:LocalProbBySumProduct})
\begin{equation} \label{eqn:plocalGenMult}
p_\text{local}(x,y\cond \hat\theta) 
  \propto g(x,y,\hat\theta) \msgf{\mu}{X}(x) \msgb{\mu}{Y}(y)
\end{equation}
with $g(x,y,\theta)$ as in Table~\ref{tab:GaussMultEM} (see also (\ref{eqn:gxythetaProofs}))
and where $\msgf{\mu}{X}$ and $\msgb{\mu}{Y}$ 
are the incoming Gaussian sum-product messages 
with parameters $\msgf{m}{X}$ and $\msgf{V}{X}$ (or $\msgf{W}{X} = \msgf{V}{X}^{-1}$)
and $\msgb{m}{Y}$ and $\msgb{V}{Y}$ (or $\msgb{W}{Y} = \msgb{V}{Y}^{-1}$),
respectively.

\begin{figure}
\begin{center}
\setlength{\unitlength}{0.85mm}
\begin{picture}(55,22.5)(0,0)
%
\put(0,2.5){\vector(1,0){14}}   \put(5,5.5){\cent{$X$}}
\put(14,-1){\framebox(7,7){$A$}}
\put(21,2.5){\vector(1,0){14}}  \put(28,5.5){\cent{$U$}}
\put(35,17.5){\framebox(5,5){}} \put(33,19){\pos{r}{$\calN(0,V_Z)$}}
\put(37.5,17.5){\vector(0,-1){12.5}}  \put(36.5,12){\pos{r}{$Z$}}
\put(35,0){\framebox(5,5){$+$}}
\put(40,2.5){\vector(1,0){15}}  \put(49,5.5){\cent{$Y$}}
\end{picture}
\vspace{2mm}
\caption{\label{fig:FactorGraphAppProosTableVxmx}%
Factor graph for Appendix~\ref{appsec:ProofsTableVxmx}.%
}
\end{center}
\end{figure}
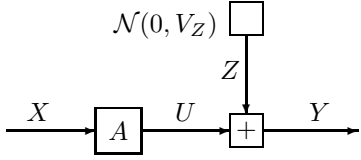

Throughout this section, $\Theta=\hat\theta$ is fixed 
and we will simply write $A$ instead of $A(\hat{\theta})$.
The factor graph of Table~\ref{tab:GaussMultEM} then reduces 
to the factor graph of \Fig{fig:FactorGraphAppProosTableVxmx}.
The desired quantities may be obtained by Gaussian sum-product message passing in 
this factor graph. 
In the following computations, we will frequently use 
Tables 2 and~3 of~\cite{LDHKLK:fgsp2007} without special notice; 
the reader is advised to have these tables at hand.

Equation (\ref{eqn:MultWX}) follows from
\begin{IEEEeqnarray}{rCl}
W_X
 & = &  \msgf{W}{X} + \msgb{W}{X} \\
 & = &  \msgf{W}{X} + A^T \msgb{W}{U} A \\
 & = &  \msgf{W}{X} + A^T \! \left( V_Z + \msgb{V}{Y} \right)^{\!-1} \! A.
        \IEEEeqnarraynumspace 
\end{IEEEeqnarray}
Equation (\ref{eqn:msgfVy}) is immediate from
\begin{IEEEeqnarray}{rCl}
\msgf{V}{Y}
 & = & \msgf{V}{U} + V_Z \\
 & = & A \msgf{V}{\!X} A^T + V_Z.
       \IEEEeqnarraynumspace
\end{IEEEeqnarray}
Equation (\ref{eqn:tildeW}) 
is the definition of $\tilde{W}$ as in \cite[eq.~(56)]{LDHKLK:fgsp2007}.
\\
Equation (\ref{eqn:VX}) follows from \cite[(I.4) and (III.8)]{LDHKLK:fgsp2007}:
\begin{IEEEeqnarray}{rCl}
V_X
 & = & \msgf{V}{X} - \msgf{V}{X} \tilde{W}_X \msgf{V}{X} \\
 & = & \msgf{V}{X} - \msgf{V}{X} A^T \tilde{W}_Y A \msgf{V}{X}.
       \IEEEeqnarraynumspace \label{eqn:ProofVX}
\end{IEEEeqnarray}
%
Equation (\ref{eqn:MultmX}) follows from
\begin{IEEEeqnarray}{rCl}
W_X m_X
 & = &  \msgf{W}{X} \msgf{m}{X} + \msgb{W}{X} \msgb{m}{X} \\
 & = &  \msgf{W}{X} \msgf{m}{X} + A^T\msgb{W}{U} \msgb{m}{U} \\
 & = &  \msgf{W}{X} \msgf{m}{X} + A^T \left( V_Z + \msgb{V}{Y} \right)^{-1} \msgb{m}{Y}.
        \IEEEeqnarraynumspace \label{eqn:ProofWXmx}
\end{IEEEeqnarray}
Using (\ref{eqn:ProofVX}) and~(\ref{eqn:ProofWXmx}), 
Equation~(\ref{eqn:MultmXalt}) follows from
\begin{IEEEeqnarray}{rCl}
m_X
 & = &  V_X W_X m_X \\
 & = &  \left( \msgf{V}{X} - \msgf{V}{X} A^T \tilde{W}_Y A \msgf{V}{X} \right) 
        \iftwocolumn{\nonumber\\ && {~~~} \cdot}{}
        \left( \msgf{W}{X} \msgf{m}{X} + A^T \left( V_Z + \msgb{V}{Y} \right)^{-1} \msgb{m}{Y} \right) 
            \IEEEeqnarraynumspace \\
 & = &  \left( I_n - \msgf{V}{X} A^T \tilde{W}_Y A \right)
        \iftwocolumn{\nonumber\\ && {~~~} \cdot}{}
        \left( \msgf{m}{X} + \msgf{V}{X} A^T \left( V_Z + \msgb{V}{Y} \right)^{-1} \msgb{m}{Y} \right).
            \IEEEeqnarraynumspace
\end{IEEEeqnarray}
Equation (\ref{eqn:MultmY}) is immediate from 
\begin{equation}
W_Y m_Y = \msgf{W}{Y} \msgf{m}{Y} + \msgb{W}{Y} \msgb{m}{Y}.
\end{equation}
Finally, Equation~(\ref{eqn:MultmYalt}), is obtained 
using \cite[(eq.~I.4)]{LDHKLK:fgsp2007}:
\begin{IEEEeqnarray}{rCl}
m_Y
 & = &  V_Y W_Y m_Y \\
 & = &  \left( \msgf{V}{Y} - \msgf{V}{Y} \tilde{W}_Y \msgf{V}{Y} \right) 
         \left( \msgf{W}{Y} \msgf{m}{Y} + \msgb{W}{Y} \msgb{m}{Y} \right) 
         \IEEEeqnarraynumspace \\
 & = &  \left( I_m - \msgf{V}{Y} \tilde{W}_Y \right)
           \left( \msgf{m}{Y} + \msgf{V}{Y} \msgb{W}{Y} \msgb{m}{Y} \right).
           \IEEEeqnarraynumspace
\end{IEEEeqnarray}

\section{Proof of (\ref{eqn:VXY})}
\label{appsec:VXY}

\begin{figure}
\begin{center}
\small
\setlength{\unitlength}{0.8mm}
\begin{picture}(95,43)(0,0)
%
%
\put(5,2.5){\vector(1,0){14}}       \put(10,5.5){\cent{$X$}}
\put(19,-1){\framebox(7,7){$A'$}}
\put(26,2.5){\vector(1,0){19}}      \put(35.5,9){\cent{$\left(\!\begin{array}{c} U\\ X \end{array}\!\right)$}}
\put(45,0){\framebox(5,5){$+$}}
 \put(45,34){\framebox(5,5){}}      \put(52,36){\pos{l}{$\calN(0,V_Z)$}}
 \put(47.5,34){\vector(0,-1){11}}   \put(46.5,28.5){\pos{r}{$Z$}}
 \put(44,16){\framebox(7,7){$B$}}
 \put(47.5,16){\vector(0,-1){11}}
\put(50,2.5){\vector(1,0){19}}      \put(59.5,9){\cent{$\left(\!\begin{array}{c} Y\\ X \end{array}\!\right)$}}
\put(69,-1){\framebox(7,7){$C$}}
\put(76,2.5){\vector(1,0){14}}      \put(85,5.5){\cent{$Y$}}
\end{picture}
\vspace{3mm}
\caption{\label{fig:MultWithOutputNoiseCorrXYFG}%
Factor graph for Appendix~\ref{appsec:VXY}.%
}
\end{center}
\end{figure}
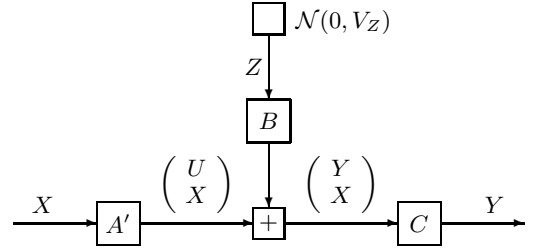

We need to compute the covariance matrix 
\begin{equation} \label{eqn:DefVXYT}
V_{XY^T} \eqdef \EE{(X-m_X)(Y-m_Y)^T}
\end{equation}
with respect to the local probability density~(\ref{eqn:plocalGenMult}). 
Consider the factor graph shown in \Fig{fig:MultWithOutputNoiseCorrXYFG}
with block matrices
\begin{IEEEeqnarray}{rCl}
A' & \eqdef & \left(\begin{array}{c} A\\ I_n \end{array}\right), \\
B  & \eqdef & \left(\begin{array}{c} I_m\\ 0 \end{array}\right), \\
C  & \eqdef & \left(\begin{array}{cc} I_m, & 0 \end{array}\right),
\end{IEEEeqnarray}
where $n$ and $m$ are the dimensions of the column vectors $X$ and $Y$, respectively.
This factor graph is obtained from the factor graph in Table~\ref{tab:GaussMultEM} 
by stretching the variable $X$ accross the adder node
so that the variables $X$ and $Y$ now appear jointly 
as components of the vector
$(Y^T, X^T)^T$
on the correspondingly labeled edge.
The closed-box function $g(x,y)$ in \Fig{fig:MultWithOutputNoiseCorrXYFG} equals 
the closed-box function $g(x,y,\hat\theta)$ in the factor graph in Table~\ref{tab:GaussMultEM}.

The desired matrix $V_{XY^T}$ is the lower left corner 
of the covariance matrix
\begin{equation}
V_{\vectsubscript{Y}{X}} 
= \left( \begin{array}{cc}
   V_Y  &  V_{X^TY} \\
   V_{XY^T} & V_X
   \end{array} \right),
   \IEEEeqnarraynumspace
\end{equation}
which can be computed by 
Gaussian sum-product message passing in \Fig{fig:MultWithOutputNoiseCorrXYFG}. 
As in Appendix~\ref{appsec:ProofsTableVxmx}, we will use 
Tables 2 and~3 of~\cite{LDHKLK:fgsp2007} without special notice. 
We have 
\begin{IEEEeqnarray}{rCl}
\msgf{V}{\vectsubscript{U}{X}}
 & = &  A' \msgf{V}{\!X} (A')^T \\
 & = &  \left(\begin{array}{cc} 
            A \msgf{V}{\!X} A^T & A \msgf{V}{\!X}\\ 
            \msgf{V}{\!X} A^T & \msgf{V}{\! X} \end{array}\right)
            \IEEEeqnarraynumspace
\end{IEEEeqnarray}
and
\begin{IEEEeqnarray}{rCl}
\msgf{V}{\vectsubscript{Y}{X}}
 & = &  \msgf{V}{\vectsubscript{U}{X}} + B V_Z B^T \\
 & = &  \left(\begin{array}{cc} 
            A \msgf{V}{\!X} A^T + V_Z  &  A \msgf{V}{\!X}\\ 
            \msgf{V}{\!X} A^T          &  \msgf{V}{\! X} \end{array}\right).
            \IEEEeqnarraynumspace
\end{IEEEeqnarray}
We also have
\begin{IEEEeqnarray}{rCl}
W_{\vectsubscript{Y}{X}}
 & = &  \msgf{W}{\vectsubscript{Y}{X}} + \msgb{W}{\vectsubscript{Y}{X}} \\
 & = &  \msgf{V}{\vectsubscript{Y}{X}}^{-1} + C^T \msgb{V}{Y}^{-1} C
           \IEEEeqnarraynumspace
\end{IEEEeqnarray}
and the Matrix Inversion Lemma 
(see, e.g., \cite[eq.~(181)]{LDHKLK:fgsp2007}) 
yields
\begin{IEEEeqnarray}{rCl}
V_{\vectsubscript{Y}{X}} 
 & = &  \msgf{V}{\vectsubscript{Y}{X}} 
           - \msgf{V}{\vectsubscript{Y}{X}} C^T 
        \iftwocolumn{\nonumber\\ && {~~~~~~~~~~}\cdot}{}
         \left( \msgb{V}{Y} + C \msgf{V}{\vectsubscript{Y}{X}} C^T \right)^{\!-1} 
                  C \msgf{V}{\vectsubscript{Y}{X}}
         \IEEEeqnarraynumspace\\
 & = &  \msgf{V}{\vectsubscript{Y}{X}} 
           - \msgf{V}{\vectsubscript{Y}{X}} C^T 
        \iftwocolumn{\nonumber\\ && {~~~~~~~~~~}\cdot}{}
         \left( \msgb{V}{Y} + A \msgf{V}{\!X} A^T + V_Z \right)^{\!-1} 
             C \msgf{V}{\vectsubscript{Y}{X}} 
             \IEEEeqnarraynumspace\\
 & = &  \msgf{V}{\vectsubscript{Y}{X}} 
           - \left(\!\begin{array}{c} A \msgf{V}{\!X} A^T + V_Z \\ \msgf{V}{\!X} A^T \end{array}\!\right)
          \iftwocolumn{\nonumber\\ && {~~~~~~~~~~~~~~~~}\cdot}{}
           \left( \msgb{V}{Y} + A\msgf{V}{\!X}A^T + V_Z \right)^{\!-1} 
         \iftwocolumn{\nonumber\\ && {~~~~~~~~~~~~~~~~}\cdot}{}
           \Big( A \msgf{V}{\!X} A^T + V_Z, \,  A \msgf{V}{\!X} \Big).
             \IEEEeqnarraynumspace
\end{IEEEeqnarray}
The lower left corner of this matrix is
\begin{IEEEeqnarray}{rCl}
V_{XY^T}
 & = &  \msgf{V}{\!X} A^T
           - \msgf{V}{\!X} A^T
             \left( \msgb{V}{Y} + A\msgf{V}{\!X}A^T + V_Z \right)^{\!-1}
          \iftwocolumn{\nonumber\\ && {~~~~~~~~~~~~~~~~~~~~~~~}\cdot}{}
            \Big( A \msgf{V}{\!X} A^T + V_Z \Big) 
            \IEEEeqnarraynumspace\\
 & = &  \msgf{V}{\!X} A^T \left( \msgb{V}{Y} + A\msgf{V}{\!X}A^T + V_Z \right)^{\!-1}
        \iftwocolumn{\nonumber\\ && {}\cdot}{}
            \Big( \left( \msgb{V}{Y} + A\msgf{V}{\!X}A^T + V_Z \right) 
        \iftwocolumn{\nonumber\\ && {~~~~~~~~~~~~~~~~~~~~}}{}
           - \Big( A \msgf{V}{\!X} A^T + V_Z \Big)
           \Big) \IEEEeqnarraynumspace \\
 &= &  \msgf{V}{\!X} A^T \left( A\msgf{V}{\!X}A^T + V_Z + \msgb{V}{Y} \right)^{\!-1}
           \msgb{V}{Y}
           \label{eqn:EMProofVXYT}
\end{IEEEeqnarray}
and using
\begin{equation}
\tilde{W}_Y = \left( A\msgf{V}{X}A^T + V_Z + \msgb{V}{Y}\right)^{-1}
\end{equation}
from (\ref{eqn:tildeW}) and (\ref{eqn:msgfVy}) yields~(\ref{eqn:VXY}).


\newcommand{\COM}{IEEE Trans.\ Communications}
\newcommand{\COMMag}{IEEE Communications Mag.}
\newcommand{\IT}{IEEE Trans.\ Information Theory}
\newcommand{\JSAC}{IEEE J.\ Select.\ Areas in Communications}
\newcommand{\SP}{IEEE Trans.\ Signal Proc.}
\newcommand{\SPMag}{IEEE Signal Proc.\ Mag.}
\newcommand{\ProcIEEE}{Proceedings of the IEEE}


\end{document}